\documentclass[11pt]{article}
\usepackage{multirow}
\usepackage{subcaption}
\usepackage{mathtools}
\usepackage{stmaryrd}
\usepackage{amsthm}
\usepackage{float}
\usepackage[width=175mm,top=20mm,bottom=25mm,bindingoffset=0mm]{geometry}
\usepackage{subcaption}

\usepackage{url} 
\usepackage[ruled,lined]{algorithm2e}
\usepackage{graphicx}
\usepackage{tikz}
\usetikzlibrary{automata, positioning, arrows}

\usepackage{amsfonts}
\usepackage{extarrows}
\usepackage{multirow}

\theoremstyle{plain}
\newtheorem{theorem}{Theorem}
\newtheorem{lemma}{Lemma}

\theoremstyle{definition}
\newtheorem{defi}{Definition}

\theoremstyle{definition}
\newtheorem{notation}{Notation}

\theoremstyle{remark}

\newcommand{\intI}{\iota^\mathrm{I}}
\newcommand{\intII}{\iota^\mathrm{I\times U}}
\newcommand{\intIII}{\iota^\mathrm{N\times U}}
\newcommand{\evoI}{\epsilon^\mathrm{I}}
\newcommand{\evoII}{\epsilon^\mathrm{N}}

\newcommand\myh{\parbox[0pt][5mm][c]{0cm}{}}
\newcommand\myhh{\parbox[0pt][4mm][c]{0cm}{}}

\newcommand{\ovun}[1]{\overline{\underline{\mathbf{#1}}}}

\newcommand\fullmoon{\hspace{1.5mm}\parbox[0pt][2.5mm][c]{0cm}{\circle{2.5mm}}\hspace{1.5mm}}

\AtBeginDocument{
  \providecommand\BibTeX{{
    \normalfont B\kern-0.5em{\scshape i\kern-0.25em b}\kern-0.8em\TeX}}}

\begin{document}
\title{Proven Distributed Memory Parallelization of Particle Methods}
\maketitle

\author{\textbf{Johannes Pahlke}\\
	Technische Universit\"at Dresden, Faculty of Computer Science, Dresden, Germany\\
	Max Planck Institute of Molecular Cell Biology and Genetics, Dresden, Germany\\
	Center for Systems Biology Dresden, Dresden Germany}\\

\author{ \textbf{Ivo F. Sbalzarini}\\
	Technische Universit\"at Dresden, Faculty of Computer Science, Dresden, Germany\\
	Max Planck Institute of Molecular Cell Biology and Genetics, Dresden, Germany\\
	Center for Systems Biology Dresden, Dresden Germany\\
	Cluster of Excellence Physics of Life, TU Dresden, Dresden, Germany\\
	Center for Scalable Data Analytics and Artificial Intelligence (ScaDS.AI) Dresden/Leipzig, Germany}

\begin{abstract}
	We provide a mathematically proven parallelization scheme for particle methods on distributed-memory computer systems. Particle methods are a versatile and widely used class of algorithms for computer simulations and numerical predictions in various applications, ranging from continuum fluid dynamics and granular flows, using methods such as Smoothed Particle Hydrodynamics (SPH) and Discrete Element Methods (DEM) to Molecular Dynamics (MD) simulations in molecular modeling. Particle methods naturally lend themselves to implementation on parallel-computing hardware. So far, however, a mathematical proof of correctness and equivalence to sequential implementations was only available for shared-memory parallelism. Here, we leverage a formal definition of the algorithmic class of particle methods to provide a proven parallelization scheme for distributed-memory computers. We prove that these parallelized particle methods on distributed memory computers are formally equivalent to their sequential counterpart for a well-defined class of particle methods. Notably, the here analyzed parallelization scheme is well-known and commonly used. Our analysis is, therefore, of immediate practical relevance to existing and new parallel software implementations of particle methods and places them on solid theoretical grounds.
\end{abstract}

\begin{keywords}
	particle methods, simulation algorithms, formal definition, algorithmics, software engineering, parallelization, distributed memory, meshfree methods
\end{keywords}

\section{Introduction}
High-performance computing (HPC) is becoming increasingly important in research,
especially in the life sciences \cite{Verma2018,Karr2012}, where many physical experiments are not feasible due to ethical reasons or technical limitations in control and observation.
At the same time, the power of computer hardware is increasing 
, mainly due to massive parallelization.
This computing power enables the simulation of increasingly complex models,
but demands elaborate code and long development times.
Therefore, generic simulation frameworks are needed to bridge the gap between accessible programming and multi-hardware parallelization.
At the foundation of these frameworks are generic and parallelizable numerical methods.

Prominent examples of such methods belong to the algorithmic class of particle methods, which has widespread use in scientific computing. Applications range from computational fluid dynamics \cite{Cottet:1990} over molecular dynamics simulations \cite{Nelson1996}
\sloppy to particle-based image processing methods \cite{Cardinale:2012, Afshar:2016}, encompassing many well-known numerical methods, such as Discrete-Element Methods (DEM)~\cite{Walther:2009}, Molecular Dynamics (MD)~\cite{Alder:1957}, Reproducing Kernel Particle Methods (RKPM)~\cite{Liu:1995}, Particle Strength Exchange (PSE)~\cite{Degond:1989a, Eldredge:2002}, Discretization-Corrected PSE (DC-PSE)~\cite{Schrader:2010, Bourantas:2016}, and Smoothed Particle Hydrodynamics (SPH)~\cite{Gingold:1977, Monaghan:2005}.
Additionally, particle methods have been efficiently parallelized on both shared, and distributed memory systems \cite{Reynders:1996a, Sbalzarini:2006b, Iwasawa:2016, Karol:2018, Incardona:2019}. 

Recently, particle methods were mathematically defined \cite{Bamme2021}, enabling their formal study independent of an application. Leveraging this formal definition, particle methods have been proven to be parallelizable on shared-memory systems under certain conditions \cite{Bamme2021}.
However, no such results have been available for the distributed-memory parallelism prevalent in HPC. It was, therefore, not known under which conditions a distributed-memory implementation of a particle method is formally equivalent to its sequential counterpart, i.e., computes the same results for any possible input. 

Here, we provide a proof of the equivalence of a distributed-memory parallelization scheme for particle methods and analyze the conditions under which it is valid. Our analysis is independent of a specific application or a specific numerical method. The proof covers a broad class of particle-based algorithms. Moreover, the parallelization scheme we analyze is well-known and commonly used in practical distributed-memory implementations. 

The considered parallelization scheme is based on the classic cell-list algorithm \cite{Hockney:1988} and a checkerboard-like domain decomposition. 
We formalize this scheme in mathematical equations as well as a Nassi-Shneiderman diagram. We then provide an exhaustive list of conditions a particle method must fulfill for the scheme to be correct. Under these conditions, we prove the equivalence of the presented parallelization scheme to the underlying sequential particle method. Furthermore, we use the presented formal analysis to infer the scheme's time complexity and parallel scalability bounds. 

We hope this work provides a starting point for the complexity bounds and correctness of distributed parallel codes in scientific HPC. In the long run, such efforts could lead to the development of provably correct software frameworks for scientific computing.

\section{Background}
We introduce the background.

\subsection{Terminology and Notation}

We introduce the notation and terminology used and define the underlying mathematical concepts.

\begin{defi} \label{def:def1}\label{def:kleenstar}
	The  \textbf{Kleene star} $A^*$ is the set of all tuples of elements of a set $A$, including the empty tuple $()$. It is defined using the Cartesian product as follows:
	\begin{align}
		A^0\!:=& \{ () \},\ 
		A^1\! := A,\
		A^{n+1}\!:= A^n\!\times\! A \ \text{ for } n \in \mathbb N_{>0}\\
		A^*:= & \bigcup_{j=0}^{\infty} A^j .
	\end{align}
	
\end{defi}

\begin{notation}
	We use \textbf{bold symbols} for tuples of arbitrary length, e.g.
	\begin{equation}
		\mathbf p \in P^*.
	\end{equation}
\end{notation}

\begin{notation}
	We use \textbf{regular symbols with subscript indices} for the elements of these tuples, e.g.
	\begin{equation}
		\mathbf p = (p_1,...,p_n).
	\end{equation}
\end{notation}

\begin{notation}
	We use \textbf{regular symbols} for tuples of determined length with specific element names, e.g.
	\begin{equation}
		p = (a,b,c)\in A\times B \times C.
	\end{equation}
\end{notation}

\begin{notation}
	We use the same \textbf{indices} for tuples of determined length and their named elements to identify them, e.g.
	\begin{equation}
		p_j = (a_j,b_j,c_j).
	\end{equation}
\end{notation}

\begin{notation}
	We use \textbf{underlined symbols} for vectors, e.g.
	\begin{equation}
		\underline{v} \in A^n.
	\end{equation}
\end{notation}

\begin{defi}
	Be $a \in \mathbb R$, $a = z+r$ with $z \in \mathbb Z$, $r \in \mathbb R$, and $0 \leq r < 1$.
	Then \textbf{rounding down} of a \textbf{real number} $a \in \mathbb R$ is defined as
	\begin{equation}
		\lfloor a\rfloor :=z.
	\end{equation}
\end{defi}

\begin{defi}
	\textbf{Rounding down of a vector} $\underline v \in \mathbb R^d$ is defined element-wise
	\begin{equation}
		\lfloor \underline{v}\rfloor :=\left(\begin{array}{ c }
			\lfloor v_{1}\rfloor \\
			\vdots \\
			\lfloor v_{d}\rfloor 
		\end{array}\right).
	\end{equation}
\end{defi}

\begin{defi}
	The \textbf{number of elements of a tuple} $\mathbf{p}=\left(p_1,...,p_n\right)\in P^*$ is defined as
	\begin{equation}
		\vert \mathbf p \vert := n.
	\end{equation}
\end{defi}

\begin{defi}\label{def:vectorlength}
	The \textbf{Euclidean-length of a vector} $\underline v \in \mathbb R^d$ is defined by the $\ell ^2$-norm
	\begin{equation}
		\left| \underline{v}\right| =\left|\begin{array}{ c }
			v_{1} \\
			\vdots \\
			v_{d} 
		\end{array}\right|
		:= \sqrt{v_1^2+\ldots+v^2_d}.
	\end{equation}
\end{defi}

\begin{defi} \label{def:compo}
	The \textbf{composition operator} $*_h$ of a binary function $h: A \times B \rightarrow A$ is recursively defined as:  
	\begin{align}
		&*_h: A\times B^*  \rightarrow A\\[6pt]
		&a *_h () := a\\
		&a *_h (b_1,b_2,...,b_n) := h(a,b_1) *_h (b_2,...,b_n).
	\end{align}
\end{defi}

\begin{defi} \label{def:concat}
	The \textbf{concatenation} $\circ : A^* \times A^* \rightarrow A^*$ of tuples $\left(a_1,...,a_n\right) , \left(b_1,..,b_m \right)\in A^*$ is defined as:
	\begin{equation}
		\left(a_1,...,a_n\right) \circ \left(b_1,..,b_m \right) 	:=  \left(a_1,...,a_n,b_1,..,b_m \right).
	\end{equation}
	The \textbf{big concatenation} $\fullmoon$ of tuples $\mathbf{a}_1,..., \mathbf{a}_n \in A^*$ is defined as:
	\begin{equation}
		\underset{j=1}{\overset{n}{\fullmoon }} \mathbf{a}_j	:=  \mathbf{a}_1\circ ...\circ \mathbf{a}_n.
	\end{equation}
\end{defi}

\begin{defi} \label{def:subtuple}
	We define the \textbf{construction of a subtuple} $\mathbf{b}~\!\in~\!A^*$ of $\mathbf a\in A^*$.
	Be $f: A^* \times \mathbb N \rightarrow \{\top, \bot \} $ ($\top = true$, $\bot = false$) the condition for an element $a_j$ of the tuple $\mathbf a$ to be in $\mathbf b$. $\mathbf b=(a_j \in \mathbf a: f(\mathbf a, j))$ defines a subtuple of $\mathbf a$ as:
	\begin{align}
		\begin{split}
			&\mathbf{ b}= (a_j \in \mathbf a: f(\mathbf a, j)) := (a_{j_1},...,a_{j_n})  \\
			\leftrightarrow \quad& \mathbf a= (a_1,..., a_{j_1},...,a_{j_2},...,a_{j_n},...,a_m)\\
			&\land \quad \forall k\in \{1,..,n\}:  \; f(\mathbf a, j_k)=\top . 
		\end{split}
	\end{align}
\end{defi}

\begin{defi}
	\sloppy
	A	\textbf{permutation} $\sigma$ is a bijective function mapping the finite set $A$ ($|A|~<~\infty$) to itself
	\begin{equation}
		\sigma: A \rightarrow A.
	\end{equation}
\end{defi}

\begin{defi}
	Be $\mathbf{a} =( a_{1} ,\ldots,a_{n}) \in A^{n}$ a tuple and $\sigma:\{a_{1} ,\ldots,a_{n}\}\rightarrow \{a_{1} ,\ldots,a_{n}\}$ a permutation. Then, the \textbf{permutation of a tuple} 
	is defined as:
	\begin{align}
		\sigma (\mathbf{a}) & :=\left( \sigma ( a_{1}) ,\ldots,\sigma ( a_{n})\right).
	\end{align}
\end{defi}

\begin{defi}
	Be $\alpha=(a_1,\ldots ,a_{n}) \in A_1\times \ldots \times A_{n}$ a tuple.
	Then, an \textbf{element} $\langle \alpha \rangle_{j}$ of a tuple is defined as
	\begin{align}
		\langle \alpha \rangle_{j}:= a_j,
	\end{align}
	and a \textbf{collection tuple} $\langle \alpha \rangle_{(j_1,\ldots,j_{m})}$ of a tuple with $j_1,\ldots,j_{m} \in \{1,\ldots,n\}$ is defined as
	\begin{align}
		\langle \alpha \rangle_{(j_1,\ldots,j_{m})}:= (a_{j_1},\ldots,a_{j_{m}}).
	\end{align}
\end{defi}

\begin{defi}
	A \textbf{subresult} ${}_{k}f$ of a function\\ 
	\begin{equation}f:A_1\times\ldots\times A_{n} \rightarrow B_1\times\ldots\times B_{m}  
	\end{equation}
	is defined as
	\begin{equation}
		{}_{k}f(a_1,\ldots,a_{n}):=\langle f(a_1,\ldots,a_{n}) \rangle_k \quad \text{with}\quad k\in \{1,\ldots,m\}.
	\end{equation}
\end{defi}

\begin{notation}
	We use a \textbf{big number with over- and underline} for a vector with the same number for all entries, e.g.
	\begin{equation}
		\overline{\underline{\mathbf{1}}}:=(1,\ldots,1)^{\mathbf{T}} \in \mathbb{N}^d.
	\end{equation}
\end{notation}

\begin{defi} \label{def:indexTransform}
	Be the dimension of the domain $d$,
	the vectorial index space 
	\begin{equation}
		\mathbb{N}^d \cap \left[\overline{\underline{\mathbf{1}}},\overline{\underline{\mathbf{I}}}\right]
		\qquad\text{with} \quad  \overline{\underline{\mathbf{I}}} =\left(I_1,\ldots,I_{d}\right)^{\mathbf{T}} \in N^d_{>0},
	\end{equation}
	and the corresponding scalar index space
	\begin{equation}
		\left\{1,\ldots, \prod _{t=1}^{d} I_{t}\right\}
	\end{equation}
	Then, the \textbf{translation of a scalar index to a vectorial index} is
	\begin{equation}
		{}^{\overline{\underline{\mathbf{I}}}} \iota : \left\{1,\ldots, \prod _{t=1}^{d} I_{t}\right\}\rightarrow \mathbb{N}^d \cap \left[\overline{\underline{\mathbf{1}}},\overline{\underline{\mathbf{I}}}\right]
	\end{equation}
	and defined as
	\begin{equation}
		{}^{\overline{\underline{\mathbf{I}}}} \iota ( j) :=\left(
		\begin{array}{ c }
			( j-1) -\left\lfloor \frac{j-1}{I_{1}}\right\rfloor I_{1} +1\\
			\left\lfloor \frac{j-1}{I_{1}}\right\rfloor -\left\lfloor \frac{j-1}{I_{1} I_{2}}\right\rfloor I_{2} +1\\
			\vdots \\
			\left\lfloor \frac{j-1}{\prod _{t=1}^{l-1} I_{t}}\right\rfloor -\left\lfloor \frac{j-1}{\prod _{t=1}^{l} I_{t}}\right\rfloor I_{l} +1\\
			\vdots \\
			\left\lfloor \frac{j-1}{\prod _{t=1}^{d-2} I_{t}}\right\rfloor -\left\lfloor \frac{j-1}{\prod _{t=1}^{d-1} I_{t}}\right\rfloor I_{d-1} +1\\
			\left\lfloor \frac{j-1}{\prod _{t=1}^{d-1} I_{t}}\right\rfloor +1
		\end{array}
		\right).
	\end{equation}
	The \textbf{backward translation of a vectorial index to a scalar index} is
	\begin{equation}
		{}^{\overline{\underline{\mathbf{I}}}} \iota ^{-1}:    \mathbb{N}^d \cap \left[\overline{\underline{\mathbf{1}}},\overline{\underline{\mathbf{I}}}\right] \rightarrow \left\{1,\ldots, \prod _{t=1}^{d} I_{t}\right\},
	\end{equation}
	and defined as
	\begin{equation}
		^{\overline{\underline{\mathbf{I}}}} \iota ^{-1} (\underline{j} ):=\ 1 + ( j_{1} -1) +( j_{2} -1) I_{1} +( j_{3} -1) I_{1} I_{2} +\ldots+( j_{l} -1)\prod _{t=1}^{l-1} I_{t} +\ldots+( j_{d} -1)\prod _{t=1}^{d-1} I_{t} .
	\end{equation}
\end{defi}

	\subsection{Mathematical Definition of Particle Methods}
	We summarize the findings of the particle methods definition paper \cite{Bamme2021}.
	The definition of particle methods defines how a particle method should be formulated. The basic idea is that particles are collections of properties, e.g., position, velocity, mass, color, etc. Each particle interacts pairwise with its neighbors to change its properties and afterward evolves on its own to further alter its properties. For convenience, there is also a global variable that is unusually a collection of particle-unspecific properties, e.g., simulation time, overall energy, etc.
	
	More formally, the definition of particle methods consists of three parts, the algorithm, the instance, and the state transition.  
	
	\textbf{Particle Method Algorithm}\label{sec:defPM:definition} \label{sec:ParticleMethodsDefinition:Algorithm}
	The definition of a particle method algorithm encapsulates the structural elements of its implementation in a small set of data structures and functions that need to be specified at the onset. 
	
	In detail, the components are defined as follows:
	\begin{defi}
		A \textbf{particle method algorithm} is a 7-tuple $(P, G, u, f, i, e, \mathring e)$, consisting of the two data structures
		\begin{align}
			\label{eq:defPM:P}
			&P  := A_1 \times A_2 \times ... \times A_n 
			&&\text{the particle space,}\\
			\label{eq:defPM:G}
			&G := B_1 \times B_2 \times ... \times B_m  
			&&\text{the global variable space,}
		\end{align}
		such that $[G\times P^*]$ \footnotemark 
		is the {\em state space} of the particle method, and five functions: 
		\begin{align}
			\label{eq:defPM:u}
			&u: [G \times P^*] \times \mathbb N \rightarrow \mathbb N^* \ \footnotemark[\value{footnote}]
			&&\text{the neighborhood function,}\\
			\label{eq:defPM:f}
			&f:  G \rightarrow \{ \top,\bot \} 
			&&\text{the stopping condition,}\\
			\label{eq:defPM:i}
			&i:  G \times P \times P \rightarrow P\times P  
			&&\text{the interact  function,}\\
			\label{eq:defPM:e}
			&e:  G \times P\rightarrow G \times P^*  \ \footnotemark[\value{footnote}]
			&&\text{the evolve function,}  \\
			\label{eq:defPM:eg}
			&\mathring{e} :  G \rightarrow G   
			&&\text{the evolve function of the global variable.}
		\end{align}
	\end{defi}
	\footnotetext{The Kleene star on $P$ is defined as $P^*:=  \bigcup_{j=0}^{\infty} P^j$ with $P^0=()$, $P^1=P$, and $P^{n+1}=P^n\times P$.}
	
	\textbf{Particle Method Instance} \label{sec:defPM:instance}
	The particle method instance describes the initial state of the particle method. Hence it relies on the data structures of the particle method algorithm.
	\begin{defi} \label{def:PMI} 
		An initial state defines a \textbf{particle method instance} for a given particle method algorithm $(P, G, u, f, i, e, \mathring e)$:
		\begin{equation}
			\label{eq:defPMI:gp}[g^1,\mathbf{p}^1] \in [G\times P^*].
		\end{equation}
		The instance consists of an initial value for the global variable $g^1 \in G$ and an initial tuple of particles $\mathbf p^1 \in P^*$.
	\end{defi}

	\textbf{Particle State Transition Function}
	The particle method state transition function describes how a particle method proceeds from the instance to the final state by using the particle method algorithm. 
	The state transition function consists of a series of state transition steps. This series ends when the stoping function $f$ returns $true$ ($\top$). The state transition function is the same for each particle method except for the specified functions of the particle method algorithm. 
	\begin{defi}
		The {\em state transition function} $S :  [G\times P^*] \rightarrow [G\times P^*]$ is defined with the following Nassi-Shneiderman diagram:
		\newcounter{counter}\stepcounter{counter}
		\newcommand{\ct}{\footnotesize{\thecounter\stepcounter{counter}}}
		\begin{figure}[H]
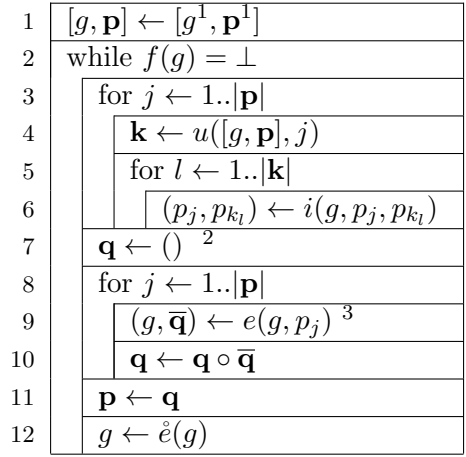

			\centering
			\label{tab:transitionStd}
			\begin{tabular}{ r| *{4}{l}|}
				\cline{2-5}  
				\ct & \multicolumn{4}{|l|}{$ [g, \mathbf p] \gets [g^1,\mathbf p^1]$  \myh}  \\  
				\cline{2-5} 
				\ct & \multicolumn{4}{|l|}{while $f(g)=\bot$ \myh}  \\ 
				\cline{3-5}
				\ct && \multicolumn{3}{|l|}{for $j \gets 1..\vert \mathbf p \vert$ \myh}  \\ 
				\cline{4-5}
				\ct && \multicolumn{1}{|l}{} & \multicolumn{2}{|l|}{$\mathbf k \gets u([g, \mathbf p], j)$ \myh} \\ 
				\cline{4-5}
				\ct&& \multicolumn{1}{|l}{} & \multicolumn{2}{|l|}{for $l \gets 1..\vert \mathbf k \vert$ \myh} \\ 
				\cline{5-5}
				\ct&& \multicolumn{1}{|l}{} & \multicolumn{1}{|l}{}& \multicolumn{1}{|l|}{$(p_j,p_{k_l}) \gets i(g,p_j, p_{k_l})$ \myh} \\ 
				\cline{3-5}
				\ct&&  \multicolumn{3}{|l|}{$\mathbf{q} \gets () \ $ \footnotemark{}  \myh}  \\ 
				\cline{3-5}
				\ct&& \multicolumn{3}{|l|}{for $j \gets 1..\vert \mathbf p \vert$ \myh}  \\ 
				\cline{4-5}
				\ct&&  \multicolumn{1}{|l|}{} &\multicolumn{2}{|l|}{$(g,\overline{\mathbf{q}} ) \gets e(g,p_j) \ $\footnotemark{} \myh}  \\
				\cline{4-5}
				\ct& & \multicolumn{1}{|l|}{} &\multicolumn{2}{|l|}{$\mathbf{q}  \gets \mathbf{q} \circ \overline{\mathbf{q}} $ \myh}  \\
				\cline{3-5}
				\ct& &\multicolumn{3}{|l|}{$\mathbf p \gets \mathbf{q}$\myh}\\
				\cline{3-5}
				\ct&&\multicolumn{3}{|l|}{$g \gets \mathring e(g)$\myh}\\
				\cline{2-5}  
			\end{tabular}
			\caption{Nassi-Shneiderman-Diagram of the state transition function $S$}
		\end{figure}
	\end{defi}
	\addtocounter{footnote}{-1}
	\footnotetext{ $\mathbf{q}$ is an intermediate result.}
	\stepcounter{footnote}
	\footnotetext{ $\overline{\mathbf{q}}$ is an intermediate result.}
	
	The Nassi-Shneiderman-Diagram corresponds to formulas that mathematically define the state transition function $S$. It is divided into sup-function for better understandably.  For each formula, we indicate which lines ({\footnotesize 1}-{\footnotesize 12}) correspond to it.
	
	All interact sup-functions return a changed particle tuple $\mathbf{p}$
	The first interact sup-function $\intI$ calculates one interaction and formalizes line {\footnotesize 6},
	\begin{multline}
		\label{eq:defPM:si1}
		\intI ([g,\mathbf p], j,k) :=(p_1,..,p_{j-1},\overline p_j, p_{j+1},...,p_{k-1},\overline p_k p_{k+1},...,p_{\vert\mathbf p\vert}),  \\
		\hspace{13mm} \text{with }\quad   (p_1,...,p_{\vert \mathbf p\vert}) = \mathbf p , \quad\left(\overline p_j, \overline p_k\right) := i\left( g,p_j,p_k\right).
	\end{multline}
	
	The second interact sup-function $\intII$ calculates the interaction of one particle with all its neighbors and formalizes line {\footnotesize 4} to {\footnotesize 6},
	\begin{equation}
		\label{eq:defPM:si2}
		\intII ([g,\mathbf p], j) := \mathbf p \; *_{\intI_{(g,j)}} \; u([g,\mathbf p], j)\
		\hspace{13mm}\text{ with } \quad \intI_{(g,j)}(\mathbf p,k):=\intI ([g,\mathbf p], j,k), 
	\end{equation}
	
	The third interact sup-function $\intII$ calculates the interaction of all particles with all their neighbors and formalizes line {\footnotesize 3} to {\footnotesize 6},
	\begin{equation}
		\label{eq:defPM:si3}
		\intIII ([g,\mathbf p]):=\mathbf p \; *_{\intII_g} (1,..,\vert \mathbf p\vert) \
		\hspace{13mm} \text{with } \quad \intII_g (\mathbf p, j):=\intII ([g,\mathbf p], j).
	\end{equation}
	
	The first evolution sup-function $\intII$ calculates the evolution of one particle and stores the result in an intermediate particle tuple $\mathbf{q}$ and in the global variable. It formalizes line {\footnotesize 9} to {\footnotesize 10},
	\begin{equation}
		\label{eq:defPM:se1}
		\evoI \big(  [g,\mathbf p], \mathbf q, j \big):= \big[ \overline g, \quad \mathbf q \circ \overline{\mathbf q}   \big]   \
		\hspace{13mm} \text{with } \quad \left(\overline g, \overline{\mathbf q}\right) := e(g,p_j).
	\end{equation}
	
	The second evolution sup-function $\intII$ calculates the evolution of all particles and returns the result in a new particle tuple $\mathbf{q}$ and global variable. It formalizes line {\footnotesize 7} to {\footnotesize 10},
	\begin{equation}
		\label{eq:defPM:se2}
		\evoII \big(  [g,\mathbf p]\big) := \big[ g, () \big] \; *_{\evoI_{\mathbf p}} \; (1,..,\vert \mathbf p\vert)
		\
		\hspace{13mm} \text{with} \quad \evoI_{\mathbf p} \big( [g,\mathbf q], j \big)  := \evoI \big( [g,\mathbf p], \mathbf q, j \big).
	\end{equation}
	
	The state transition step $s$ brings all sup-functions together and  formalizes line {\footnotesize 3} to {\footnotesize 12},
	\begin{equation}
		\label{eq:defPM:ss}
		s \left(  [g, \mathbf p] \right):=
		\big[ \mathring{e}( \overline{ g} ),\; \overline{\mathbf p}   \big]
		\qquad \text{with} 
		\quad  \left[\overline g, \overline{\mathbf p}\right] := \evoII \big(  [g,\; \intIII ([g,\mathbf p])]\big).
	\end{equation}
	
	Finally, the state transition function $S$ advances the instance to the final state by formalizing line {\footnotesize 1} to {\footnotesize 12},
	\begin{multline}\label{eq:stateTransitionFunctionDefinition}
		S([ g^1, \mathbf p^1]) = [ g^T, \mathbf p^T]  
		\qquad\longleftrightarrow \\
		f(g^T)=\top \quad\
		\land \quad \forall t\in \{2,...,T\}:\ [g^{t},\mathbf{p}^{t}]=s\left([g^{t-1},\mathbf{p}^{t-1}]\right)\ \land \ f(g^{t-1})=\bot.
	\end{multline}

\section{Distributed Pull Particle Methods without Global Operations} While most parallelization strategies tend to be algorithm-specific, the formal definition of particle methods \cite{Bamme2021} enables us to formulate, prove, and analyze a parallelization scheme for an entire class of algorithms. We consider the parallelization of particle methods on distributed-memory computers. The resulting scheme is valid for particle methods with pull interactions and without global operations. It is based on cell lists and checkerboard-like selection, a well-known approach often used in practical code implementations. Even so, the specific checkerboard-like selection used here is not usually applied. We chose it for its simplicity.

Formally, cell lists require that each particle $p$ has a position $\underline x$
\begin{equation}
	p=( \ldots,\underline{x} ,\ldots) \in P,
\end{equation}
and that the neighborhood function $u$ is based on a cutoff radius $r_c$ and not directly on indices. This makes it, in some sense, order-independent. Hence, it is restricted to the form
\begin{equation}\label{eq:constraint:neighborhood}
	u\left([ g,\mathbf{p}] ,j\right) :=\left( k\in ( 1,\ldots,|\mathbf{p} |) :p_{k} ,p_{j} \in \mathbf{p} \ \land |\underline{x}_{k} -\underline{x}_{j} |\leq r_{c} \land \Omega ( g,p_{k} ,p_{j})\right)\quad
	\text{with } \Omega :G\times P\times P\rightarrow \{\top ,\bot \},
\end{equation}
where $Omega$ is an additional constraint for more flexibility.
The position $\underline{x}$ of a particle is limited to a spatial domain of dimension $d$ for all states/times $t$
\begin{equation}\label{eq:condition:particleStayInDomain}
	\forall t\in \{1,\ldots,T\} \ \forall p_{j}^{t} \in \mathbf{p}^{t} : \underline{x}_{j}^{t} \in [\underline D_{min} ,\underline D_{max} )\subset \mathbb{R}^{d}.
\end{equation}
This is usually achieved with some boundary conditions, e.g., periodic boundary conditions. 
In addition, the position is not allowed to change by more than a cutoff radius $r_c$ in a single state transition step
\begin{equation}\label{eq:condition:MovmentIsSmallerThenRc}
	\forall t\in \{1,\ldots,T-1\} \ \forall p_{j}^{t} \in \mathbf{p}^{t} : |\underline{x}_{j}^{t} -\underline{x}_{j}^{t+1} |\leq r_{c},
\end{equation}
This has, in many cases, no impact since the traveling distance is already stronger limited to get stable simulations or correct results, e.g., MD, PSE, DEM.
The interaction is limited to a pull interaction
\begin{equation} \label{eq:condition:pullInteraction}
	\forall p_j,p_k \in P, g \in G: \,i(g,p_j, p_k)=(\overline p_j, p_k),
\end{equation}
independent of previous interactions of the interaction partner to avoid chain dependencies of interaction results
\begin{equation}\label{eq:condition:interactionPreviousInteractionIndependence}
	\forall p_{j} ,p_{k} ,p_{k'} \in P,g\in G:\ {}_{1} i_{g} (p_{j} ,{}_{1} i_{g} (p_{k} ,p_{k'} ))={}_{1} i_{g} (p_{j} ,p_{k} )\quad
	\text{with} \ \ \ \ {}_{1} i_{g} (p_{j} ,p_{k} ):=\langle i(g,p_{j} ,p_{k} )\rangle _{1},
\end{equation}
and be order-independent
\begin{equation}\label{eq:constraint:orderindependence}
	\forall p_{j} ,p_{k} ,p_{k'} \in P,g\in G:\ {}_{1} i_{g} ({}_{1} i_{g} (p_{j} ,p_{k} ),p_{k'} )={}_{1} i_{g} ({}_{1} i_{g} (p_{j} ,p_{k'} ),p_{k} ).
\end{equation}
The order independence avoids the constant resorting of the particles.  
The neighborhood function $u$ needs to be independent of previous interactions, so the cell list can be prepared before the interactions happen. Here, $*_{\intI}$ is the
composition operator defined in def.~\ref{def:compo} and used on $\intI$, whereas $\intI$ is the transition sub-function defined in eq.~\ref{eq:defPM:si1}.
\begin{equation}\label{eq:condition:neighborhoodPreviousInteractionIndependence}
	\forall j,k',k'' \in \{1,\ldots,\vert \mathbf p \vert\}, [g,\mathbf p]\in [G\times P^*]: \
	u([g,\mathbf p], j)= u([g,\mathbf p *_{\intI_{(g,k')}} (k'')], j).
\end{equation}
The evolve function $e$ does not change the global variable $g$ 
\begin{equation}\label{eq:condition:evolutionDoesntChangeGlobalVariable}
	\forall g\in G, p \in P: e(g,p)=(g,\mathbf q),
\end{equation}
to avoid global operations and their $\mathcal O \left( log(number\ of\ processes)\right)$ time complexity. 

Under these constraints on a particle method, we formulate the parallelization of any particle method onto multiple (distributed-memory) processes. A ``process'' is thereby an independent and self-contained unit of calculation. The parallelization scheme is based on the classic cell-list algorithm~\cite{Hockney:1988}. Cell lists require the definition of a cutoff radius $r_c$ to divide the domain into equal-sized Cartesian cells. The number of cells along each dimension of the computational domain is represented by the vector
\begin{equation}\label{eq:definitionOfI}
	\ovun{I}=\left(\begin{array}{ c }
		I_{1}\\
		\vdots \\
		I_{d}
	\end{array}\right)
	:=\left\lfloor \tfrac{1}{r_{c}} (\underline{D}_{\max} -\underline{D}_{\min} )+\ovun{1}\right\rfloor .
\end{equation}
The total number of cells is
\begin{equation}
	N_{\text{cell}}:=	\prod\limits_{l=1}^{d}I_l.
\end{equation}

The highest degree of parallelism is reached when assigning each cell-list cell to a separate process. Each process then has to exchange information with its direct (face-, edge-, and corner-connected) neighbors in the cell-list grid. 
We assume that each process can only communicate with one other process simultaneously and that the network behaves like a fully connected one. Then the communications between processes are subject to the risk of overlapping, potentially leading to processes having to wait for one another. Hence, one should avoid simultaneous communication with identical partners. 
Then, the simplest way to avoid overlapping communications is if all communicating processes have at least two inactive processes between them. This results in a checkerboard-like pattern.
The number of active cells in each dimension is given by the vector
\begin{equation}\label{eq:definition:kIstar}
	{}^k\ovun{I}^{*} =
	\left(\begin{array}{ c }
		{}^k	I^{*}_{1}\\
		\vdots \\
		{}^k	I^{*}_{d}
	\end{array}\right)
	:=\left\lfloor \tfrac{1}{3}\left(\ovun{I} -\ {}^{\ovun{3}} \iota (k)+\ovun{3}\right)\right\rfloor.
\end{equation}
We have then $3^d$ different communicating process configurations $k\in \{1,\ldots,3^d\}$. 
In total, the number of communicating processes for each $k$ is
\begin{equation}
	{}^{k}N^*_{\text{cell}}:=	\prod\limits_{l=1}^{d}{}^k	I^{*}_l.
\end{equation}
We address the communicating processes for each $k$ by
\begin{equation}\label{eq:definition:gamma}
	\gamma ( k,j) :=\ {}^{\ovun{I}} \iota ^{-1}\left( \ {}^{{}^k\ovun{I}^{*}}\!\! \iota (j)\cdot 3+{}^{\ovun{3}} \iota (k)-\ovun{3}\right).
\end{equation}
The $l$-th neighbor cell of the $t$-th process is 
\begin{equation}\label{eq:definition:beta}
	\beta ( t,l) :=\ {}^{\ovun I} \iota ^{-1}\left(  {}^{\ovun{I}} \iota (t)+{}^{\ovun{3}} \iota (l)-\ovun{2}\right).
\end{equation}
Note that the result of $\beta$ is undefined unless the result of ${}^{\ovun I} \iota ^{-1}$ (def. \ref{def:indexTransform}, eq. \ref{eq:definitionOfI}) is defined.\\[12pt]

We formulate the standard procedure of distributing particles into a cell list as a condition for the initial particle distribution. Hence, 
each cell $k$ initially contains a tuple of particles $\mathbf p^1_k$.
To ensure no particle is lost, we require that the concatenation of these particle tuples is a permutation $\pi$ of the initial particle tuple $\mathbf p^1$
\begin{equation}\label{eq:initialParticlesStorages:permutation}
	\mathbf{p}^1_1 \circ \ldots \circ \mathbf{p}^1_{N_{cell}}=\pi\left( \mathbf{p}^1\right)
\end{equation}
and that particles are distributed according to their position
\begin{equation}\label{eq:initialCell:ParticleDistribution}
	\forall p^1_j\in \mathbf p^1: p^1_j \in \mathbf{p}^1_{w}\ \quad \text{with }\quad w= {}^{\ovun{I}}\iota^{-1}\left(\left \lfloor \frac{1}{r_c} ( \underline{x}^1_j-\underline{D}_{\min})\right \rfloor + \ovun{1} \right).
\end{equation}
Each process has its own memory address space, containing its global variable storage $g$ and its particle storage $\underline{\mathbf p}$. The latter stores the particles within the cell-list cell assigned to that process and copies of the particles from neighboring cells. Therefore, the particle storage of each process is compartmentalized cell-wise:
\begin{equation}
	\underline{\mathbf{p}} = 
	\left( \begin{array}{ c }
		\left< \underline{\mathbf{p}}\right>_{1}\\
		\vdots\\
		\left< \underline{\mathbf{p}}\right>_{3^d}
	\end{array}\right)^{\!\!\top}\in (P^*)^{3^d},
\end{equation}
where the center entry (i.e., location $\tfrac{3^d+1}{2}$) contains the ``real'' particles of the cell assigned to that process. The other entries contain the copies of the cells from the neighboring processes in the order corresponding to the position in the cell list. 
Two processes are neighbors if the corresponding cells are direct (face-, edge-, and corner-connected) neighbors in the cell-list grid.
\begin{figure}
	\centerline{\includegraphics[width=5in]{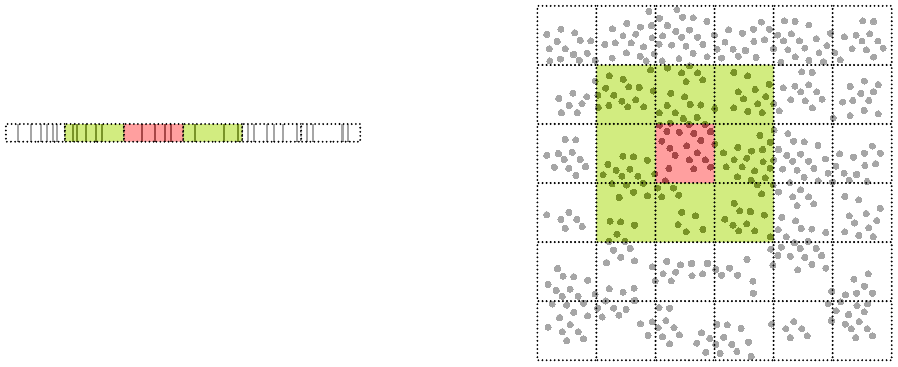}}
	\caption{Example for a cell list for one dimension (left) and two dimensions (right). The particles in one dimension are marked as lines and in two dimensions as dots. One process's particle storage overlays the cell list in green and red. The red cell marks the corresponding cell for that storage and its center storage compartment.  \label{fig:distributed:cell-list-storage}}
\end{figure}

The particle storages of all processes (PROC) together is
\begin{equation}
	\mathcal P= 
	\left(\begin{array}{ c }
		{}^{[\text{PROC}1]}\underline{\mathbf{p}} \\
		\vdots\\
		{}^{[\text{PROC}N_{\text{cell}}]}\underline{\mathbf{p}} \\
	\end{array}\right)^{\!\!\top}\in \left((P^*)^{3^d}\right)^{N_{\text{cell}}}.
\end{equation}
The initial particles tuples in each process's particle storage are
\begin{equation}
	{}^{[\text{PROC}n]}\underline{\mathbf{p}}^1 = 
	\Big( (),\ldots,(),\underbrace{\mathbf{p}^1_n}_{\tfrac{3^d+1}{2}\text{-th entry}},(),\ldots,()\Big).
\end{equation}
Then, the initial particle storage of all processes is
\begin{equation}\label{eq:initialParticlesStorages:AllProcessors}
	\mathcal{P}^1:=\left(\begin{array}{ c }
		{}^{[\text{PROC}1]}\underline{\mathbf{p}}^1 \\
		\vdots\\
		{}^{[\text{PROC}N_{\text{cell}}]}\underline{\mathbf{p}}^1 \\
	\end{array}\right)^{\!\!\top}.
\end{equation}
The global variable storage of all processes is
\begin{equation}\label{eq:initialGlobalVariableStorages:All}
	\mathcal G^1:=(g^1,\ldots,g^1)\in G^{N_{\text{cell}}}.
\end{equation}

We introduce a second global variable $\tilde{g}$ that contains the cell-list specific parameters (eq.~\ref{eq:condition:particleStayInDomain},~\ref{eq:definitionOfI})
\begin{equation}
	\tilde{g} =\left(\underline{D}_{\min} ,\underline{D}_{\max} ,d,\overline{\underline{\mathbf{I}}}\right).
\end{equation}
The function $copy_{(\tilde{g},\mathcal{P})}$ copies the center particle storage compartment of a process $\left[\text{PROC} \beta ( w,l)\right]$  to the specified compartment $l$ in the storage $\underline{\mathbf{p}}$ of another process: 
\begin{equation} \label{eq:copy:fromOneCPU}
	copy_{\left( \tilde{g} ,\mathcal P ,w\right)}(\underline{\mathbf{p}} ,l)  :=\left(\left< \underline{\mathbf{p}}\right> _{1} ,\ldots,\left< \underline{\mathbf{p}}\right> _{l-1} ,\ ^{\left[\text{PROC} \beta ( w,l)\right]}\left< \underline{\mathbf{p}}\right> _{\tfrac{3^{d} +1}{2}} ,\left< \underline{\mathbf{p}}\right> _{l+1} ,\ldots,\left< \underline{\mathbf{p}}\right> _{3^{d}}\right).
\end{equation}
This coping is done for every $k$-th cell in the checkerboard-like configuration by 
{\begin{equation}
		copy_{\tilde{g}}^{active}(\mathcal{P} ,k)  :=\left(\begin{array}{ c }
			^{\left[\text{PROC} 1\right]}\left< \underline{\mathbf{p}}\right> \\
			\vdots \\
			^{\left[\text{PROC} \gamma ( k,1) -1\right]}\left< \underline{\mathbf{p}}\right> \\
			^{\left[\text{PROC} \gamma ( k,1)\right]}\left< \underline{\mathbf{p}}\right> *_{copy_{\left(\tilde{g} ,\mathcal{P} ,\gamma ( k,1)\right)}{}_{\myhh}}\left( 1,\ldots,3^{d}\right)\\
			^{\left[\text{PROC} \gamma ( k,1) +1\right]}\left< \underline{\mathbf{p}}\right> \\
			\vdots \\
			^{\left[\text{PROC} \gamma ( k,l) -1\right]}\left< \underline{\mathbf{p}}\right> \\
			^{\left[\text{PROC} \gamma ( k,l)\right]}\left< \underline{\mathbf{p}}\right> *_{copy_{\left(\tilde{g} ,\mathcal{P} ,\gamma ( k,l)\right)}{}_{\myhh}}\left( 1,\ldots,3^{d}\right)\\
			^{\left[\text{PROC} \gamma ( k,l) +1\right]}\left< \underline{\mathbf{p}}\right> \\
			\vdots \\
			^{\left[\text{PROC} \gamma \left( k,{}^{k}N^*_{cell} \right) -1\right]}\left< \underline{\mathbf{p}}\right> \\
			^{\left[\text{PROC} \gamma \left( k,{}^{k}N^*_{cell} \right)\right]}\left< \underline{\mathbf{p}}\right> *_{copy_{\left(\tilde{g} ,\mathcal{P} ,\gamma \left( k,{}^{k}N^*_{cell} \right)\right)}{}_{\myhh}}\left( 1,\ldots,3^{d}\right)\\
			^{\left[\text{PROC} \gamma \left( k,{}^{k}N^*_{\text{cell}} \right) +1\right]}\left< \underline{\mathbf{p}}\right> \\
			\vdots \\
			^{\left[\text{PROC}N_{\text{cell}}\right]}\left< \underline{\mathbf{p}}\right> 
		\end{array}\right)^{\!\!\top}.
\end{equation}}
Doing so  for all distinct checkerboard-like configurations results in
\begin{equation}
	copy_{\tilde{g}}^{ALL}(\mathcal{P})  :=\mathcal{P} *_{copy^{active}_{\tilde{g}}}\left( 1,\ldots,3^{d} \right).
\end{equation}
After the function $copy_{\tilde{g}}^{ALL}$, each process has (copies of) all particles required to calculate the interactions and evolutions of the particles in its cell without any further inter-process communication.
We define for this scheme the interaction of all particles with their respective interaction partners by the function
\begin{equation}
	interaction_{g}(\underline{\mathbf{p}}) :=\left(\begin{array}{ c }
		\mathbf{\left< q\right> }_{1} \ *_{_{1} i_{g}}\left< \underset{w=1}{\overset{3^{d}}{\fullmoon }} \langle \underline{\mathbf{p}} \rangle _{w}\right> _{u\left(\left[ g,\underset{w=1}{\overset{3^{d}}{\fullmoon }} \langle \underline{\mathbf{p}} \rangle _{w}\right] ,z+1\right)}\\
		\vdots \\
		\mathbf{\left< q\right> }_{|\mathbf{q}| } \ *_{_{1} i_{g}}\left< \underset{w=1}{\overset{3^{d}}{\fullmoon }} \langle \underline{\mathbf{p}} \rangle _{w}\right> _{u\left(\left[ g,\underset{w=1}{\overset{3^{d}}{\fullmoon }} \langle \underline{\mathbf{p}} \rangle _{w}\right] ,z+|\mathbf{q} |\right)}
	\end{array}\right)^{\!\!\top},
\end{equation}\label{eq:stepAll:interaction}
where $z=\left| \underset{w=1}{\overset{\tfrac{3^{d}+1}{2} -1}{\fullmoon }} \langle \underline{\mathbf{p}} \rangle _{w}\right| $ 
and $\mathbf{q} =\left< \underline{\mathbf{p}}\right> _{\tfrac{3^{d} +1}{2}}$.
The step function $step_{\left(\tilde{g},g\right)}$ uses the function $interaction_{g}$ to compute the state-transition step (mostly simulation time step) including the evolutions of the particle properties and positions:
\begin{equation}\label{eq:stepAll:step}
	step_{\left(\tilde{g} ,g\right)}(\underline{\mathbf{p}})  :=\left(\begin{array}{ c }
		\left< \underline{\mathbf{p}}\right> _{1}\\
		\vdots \\
		\left< \underline{\mathbf{p}}\right> _{\tfrac{3^{d} +1}{2} -1}\\
		{}_{2}\evoII\left(\left[g,interaction_{g}\left(\underline{\mathbf{p}}\right)\right]\right)\\
		\left< \underline{\mathbf{p}}\right> _{\tfrac{3^{d} +1}{2} +1}\\
		\vdots \\
		\left< \underline{\mathbf{p}}\right> _{3^{d}}
	\end{array}\right)^{\!\!\top}.
\end{equation}
Calculating the state transition step on all processes results in
\begin{equation}\label{eq:stepAll:stepAll}
	step_{\left(\tilde{g} ,\mathcal{G}\right)}^{ALL}(\mathcal{P}) :=\left( step_{\left(\tilde{g} ,\ {}^{\left[\text{PROC} 1\right]} g\right)}\left( {}^{\left[\text{PROC} 1\right]}\underline{\mathbf{p}}\right) ,\ \ldots,\ step_{\left(\tilde{g} ,\ ^{\left[\text{PROC} |\mathcal{P} |\right]} g\right)}\left(  {}^{\left[\text{PROC} |\mathcal{P} |\right]}\underline{\mathbf{p}}\right)\right),
\end{equation}
where ${}^{\left[\text{PROC} k\right]} g:=\mathcal{\left< G\right> }_{k}$ 
and ${}^{\left[\text{PROC} k\right]}\underline{\mathbf{p}} :=\left< \mathcal{P}\right> _{k}$.

After the function $step_{\left(\tilde{g},\mathcal{G}\right)}^{ALL}$, all particles in the center storage compartments of all processes have the correct positions and properties. However, they may be in the wrong cell/ storage compartment after moving. Therefore, the particles in the center storage compartments must be re-assigned into the compartments and communicated to the new process if they have moved to another cell/ compartment.
Cell/ compartment assignment of each particle $q$ is done by
\begin{equation}\label{eq:distAll:dist}
	dist_{\left(\tilde{g} ,g,j\right)} (\underline{\mathbf{p}} ,q):=\left(\begin{array}{ c }
		\left< \underline{\mathbf{p}}  \right>_{1}\\
		\vdots \\
		\left< \underline{\mathbf{p}}   \right>_{\alpha -1}\\
		\left< \underline{\mathbf{p}}  \right >_{\alpha } \circ (q)\\
		\left< \underline{\mathbf{p}}   \right>_{\alpha +1}\\
		\vdots \\
		\left< \underline{\mathbf{p}}   \right>_{3^{d}}
	\end{array}\right)^{\!\!\top},
\end{equation}
where 
\begin{equation}
	q=( \ldots,\underline{x} ,\ldots) \in P\, ,\quad \alpha :=  {}^{\ovun 3} \iota^{-1}\left(\left\lfloor \tfrac{1}{r_{c}}(\underline{x} -\underline{D}_{min})\right\rfloor - {}^{\ovun I} \iota ( j) +\ovun 3 \right).
\end{equation}
For all particles in a center storage compartment $\mathbf{q}$, redistribution is done by the function 
\begin{equation}\label{eq:distAll:distN}
	dist_{\left(\tilde{g} ,g,j\right)}^{N}(\mathbf{q})  :=\left(() ,\ldots,()\right) *_{dist_{\left(\tilde{g} ,g,j\right)}}\mathbf{q},
\end{equation}
where $\left((),\ldots,()\right)$ is the tuple of $3^d$ empty tuples, representing the empty particle storage of a process. For all processes, the redistribution procedure is 
\begin{equation}\label{eq:distAll:distAll}
	dist_{\left(\tilde{g} ,\mathcal{G}\right)}^{ALL} (\mathcal{P} ):=\left(\begin{array}{ c }
		dist_{\left(\tilde{g} ,\ ^{\left[\text{PROC} 1\right]} g,1\right)}^{N}\left( \ ^{\left[\text{PROC} 1\right]} \left< \underline{\mathbf{p}}  \right>_{\tfrac{3^{d} +1}{2}}\right)\\
		\vdots \\
		dist_{\left(\tilde{g} ,\ ^{\left[\text{PROC} |\mathcal{P} |\right]} g,|\mathcal{P} |\right)}^{N}\left( \ ^{\left[\text{PROC} |\mathcal{P} |\right]} \left< \underline{\mathbf{p}}  \right>_{\tfrac{3^{d} +1}{2}}\right)
	\end{array}\right)^{\!\!\top},
\end{equation}
where ${}^{\left[\text{PROC} k\right]} g:=\mathcal{\left< G\right> }_{k}$ 
and ${}^{\left[\text{PROC} k\right]}\underline{\mathbf{p}} :=\left< \mathcal{P}\right> _{k}$.
After the function $dist_{\left(\tilde{g},\mathcal{G}\right)}^{ALL}$ the particles on all processes are correctly assigned to their respective storage compartments/cell-list cells, but those particles may belong to another process now, except for the particles in the center storage compartment. Therefore, communication between processes is required for the particles that have moved to a different storage compartment/cell. Collecting the particles that moved from the $\beta(w,l)$-th process to the $w$-th process is done by the function
\begin{equation}\label{eq:collectAll:collect}
	collect_{\left(\tilde{g} ,\mathcal{P} ,w\right)} (\underline{\mathbf{p}} ,l):=\left(\begin{array}{ c }
		\left< \underline{\mathbf{p}}  \right>_{1}\\
		\vdots \\
		\left< \underline{\mathbf{p}}  \right>_{\tfrac{3^{d} +1}{2} -1}\\
		\left< \underline{\mathbf{p}}  \right>_{\tfrac{3^{d} +1}{2}} \circ^{\myh} \ {}^{\left[\text{PROC} \beta (w,l)\right]} \left< \underline{\mathbf{p}}  \right>_{{}^{\ovun 3} \iota^{-1}\left(\ovun 4 -{}^{\ovun 3} \iota ( l)\right) }\\
		\left< \underline{\mathbf{p}}  \right>_{\tfrac{3^{d} +1}{2} +1}\\
		\vdots \\
		\left< \underline{\mathbf{p}}  \right>_{3^{d}}
	\end{array}\right).
\end{equation}
All processes collecting particles from the other processes simultaneously could lead to overlapping communications, hence, to race conditions or serialization.
To prevent this, the collection procedure follows the checkerboard-like pattern again. For the $k$-th checkerboard-like pattern, the collection function is
\begin{equation}\label{eq:collectAll:collectN}
	collect_{\tilde{g}}^{N} (\mathcal{P} ,k):=\left(\begin{array}{ c }
		^{\left[\text{PROC} 1\right]} \left< \underline{\mathbf{p}}  \right>\\
		\vdots \\
		^{\left[\text{PROC} \gamma (k,1)-1\right]} \left< \underline{\mathbf{p}}  \right>\\
		^{\left[\text{PROC} \gamma (k,1)\right]} \left< \underline{\mathbf{p}}  \right>*_{collect_{\left(\tilde{g} ,\mathcal{P} ,\gamma (k,1)\right)}}\left( 1,\ldots,3^{d}\right)\\
		^{\left[\text{PROC} \gamma (k,1)+1\right]} \left< \underline{\mathbf{p}}  \right>^{\myh}\\
		\vdots \\
		^{\left[\text{PROC} \gamma (k,l)-1\right]} \left< \underline{\mathbf{p}}  \right>\\
		^{\left[\text{PROC} \gamma (k,l)\right]} \left< \underline{\mathbf{p}}  \right>*_{collect_{\left(\tilde{g} ,\mathcal{P} ,\gamma (k,l)\right)}}\left( 1,\ldots,3^{d}\right)\\
		^{\left[\text{PROC} \gamma (k,l)+1\right]} \left< \underline{\mathbf{p}}  \right>^{\myh}\\
		\vdots \\
		^{\left[\text{PROC} \gamma \left( k,{}^{k}N^*_{cell}\right) -1\right]} \left< \underline{\mathbf{p}}  \right>\\
		^{\left[\text{PROC} \gamma \left( k,{}^{k}N^*_{cell}\right)\right]} \left< \underline{\mathbf{p}}  \right>*_{collect_{\left(\tilde{g} ,\mathcal{P} ,\gamma \left( k,{}^{k}N^*_{cell}\right)\right)}}\left( 1,\ldots,3^{d}\right)\\
		^{\left[\text{PROC} \gamma \left( k, {}^{k}N^*_{cell}\right) +1\right]} \left< \underline{\mathbf{p}}  \right>^{\myh}\\
		\vdots \\
		^{\left[\text{PROC} |\mathcal{P} |\right]} \left< \underline{\mathbf{p}}  \right>
	\end{array}\right)^{\!\!\top},
\end{equation}
where ${}^{\left[\text{PROC} k\right]}\underline{\mathbf{p}} :=\left< \mathcal{P}\right>_{k}$. 
The complete collection procedure is serial for the $3^d$ checkerboard-like patterns. Hence, it takes $3^d$ steps to finish. The complete collection is
\begin{equation}\label{eq:collectAll:collectAll}
	collect_{\tilde{g}}^{ALL}(\mathcal{P})  :=\mathcal{P} *_{collect_{\tilde{g}}^{N}}\left( 1,\ldots,3^{d}\right),
\end{equation}
which results in each central particle storage compartment of all processes containing the correct particles. The copies from the neighboring processes will then be updated in the subsequent iteration, again by the function $copy_{\tilde{g}}^{ALL}$.

Taking all functions together, the parallelized state-transition step for a distributed-memory particle method can be formally expressed as:
\begin{equation}\label{eq:distributed:stateTransitionStep}
	\tilde{s}_{\tilde{g}}\left([\mathcal{G} ,\mathcal{P}]\right) =\left[\left(\begin{array}{ c }
		\overset{\circ }{e}\left(\left< \mathcal{G}\right> _{1}\right)\\
		\vdots \\
		\overset{\circ }{e}\left(\left< \mathcal{G}\right> _{N_{cell}}\right)
	\end{array}\right)^{\mathbf{T}} ,collect_{\tilde{g}}^{ALL}\left( dist_{\left(\tilde{g} ,\mathcal{G}\right)}^{ALL}\left( step_{\mathcal{G}}^{ALL}\left( copy_{\tilde{g}}^{ALL} (\mathcal{P} )\right)\right)\right)\right].
\end{equation}
We use this parallel state-transition step $\tilde{s}_{\tilde{g}}$ to define the parallel state transition function similar to equation~\ref{eq:stateTransitionFunctionDefinition}
\begin{multline}\label{eq:distributed:stateTransitionFunction}
	\tilde{S}\left(\left[\mathcal{G}^{1} ,\mathcal{P}^{1}\right]\right) :=\left[\mathcal{G}^{T} ,\mathcal{P}^{T}\right]\quad  \longleftrightarrow
	\\
	f\left(\left< \mathcal{G}^{T}\right> _{1}\right) =\top \ \ \
	\land \ \ \forall t\in \{2,\ldots,T\} :\ \left[\mathcal{G}^{t} ,\mathcal{P}^{t}\right] =\tilde{s}_{\tilde{g}} \left(\left[\mathcal{G}^{t-1} ,\mathcal{P}^{t-1}\right]\right) \ \land \ \ f\left( \ \left< \mathcal{G}^{t-1}\right> _{1}\right) =\bot. 
\end{multline}

\begin{figure}[H]
	\includegraphics[width=\textwidth]{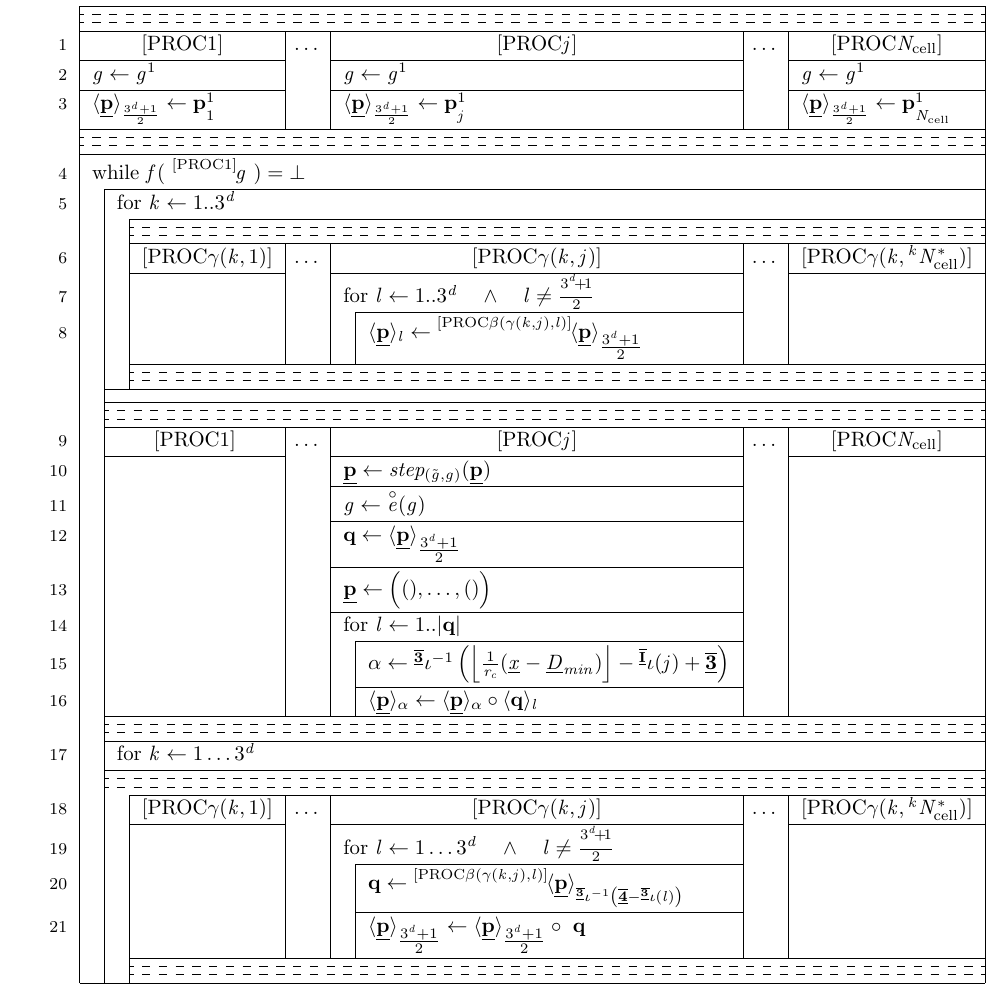}
	\caption{Nassi-Shneiderman diagram of the distributed-memory parallelization of the outer loop of a particle method with pull interaction. The dashed double lines mark parallel sections of the algorithm.}\label{fig:parallelInnerPull}
\end{figure}

The present parallelization scheme is summarized by the Nassi-Shneiderman diagram in Fig.~\ref{fig:parallelInnerPull}. 

The algorithm starts by initializing the global variable storage (line {\footnotesize 1}) and the particle storage (line {\footnotesize 3}). The global variable storage is filled with the initial global variable and the center particle storage compartment
is filled with the particles from the corresponding cell-list cell. 
The remaining particle storage compartments are filled with copies of the particles from directly adjacent neighboring cells by copying them from the respective process (lines {\footnotesize 5}-{\footnotesize 8}), which is repeated for each simulation state (line {\footnotesize 4}).
Then, each process evaluates the state-transition step of the particle method (lines {\footnotesize 10}, {\footnotesize 11}), which is only guaranteed to return the correct result for the particles of the center particle storage compartment. 
The results of the remaining particles may be corrupted by missing interactions with particles outside the process's storage.
Therefore, the algorithm proceeds to store the particles of the center storage compartment in $\mathbf{q}$ (line {\footnotesize 12}) and deletes all particles in the particle storage $\underline{\mathbf{p}}$ (line {\footnotesize 13}). 
The particles of the center storage compartment, now in $\mathbf{q}$, are redistributed to the process's particle storage compartments according to their new position (lines {\footnotesize 14}-{\footnotesize 16}).
Finally, the algorithm collects for each process all particles that are newly belonging to it and which are in the corresponding particle storage compartments on other processes (lines {\footnotesize 17}- {\footnotesize 21}).

\subsection{Lemmata}
We aim to formally prove that the above distributed-memory parallelization of a particle method is equivalent to the original sequential algorithm. To prepare the proof, we first derive a couple of useful lemmata. The proofs for all lemmata can be found in the appendix \ref{sup:lemmata}.

\begin{lemma}\label{lemma:indexTransformIsBijectiv}
	${}^{\overline{\underline{\mathbf{I}}}} \iota$ and ${}^{\overline{\underline{\mathbf{I}}}} \iota ^{-1}$ are bijections and mutual functional inverses, i.e., ${}(^{\overline{\underline{\mathbf{I}}}} \iota ^{-1})^{-1} = {}^{\overline{\underline{\mathbf{I}}}} \iota$.
\end{lemma}

\begin{lemma}\label{lemma:OrderIdependenceOfInteractSeries}
	Order independence of a series of interactions follows from the order independence of the interact function:
	\begin{align}
		& _{1} i_{g}(_{1} i_{g} (p_{j} ,p_{k} ),p_{k'}) =_{1} i_{g} (_{1} i_{g} (p_{j} ,p_{k'} ),p_{k''} )\\
		\rightarrow  & \tilde{p} \ *_{_{1} i_{g}} \ \sigma ( p_{1} ,\ldots,p_{n}) =\tilde{p} \ *_{_{1} i_{g}} \ ( p_{1} ,\ldots,p_{n}).
	\end{align}
\end{lemma}

\begin{lemma}\label{lemma:copyAll:NoReadConflicts}
	We need to poof that the function $copy_{\tilde{g}}^{ALL}$ does not induce overlapping communications.
	
	Overlapping lapping means that two processes communicate either to the same other third process or to each other.
\end{lemma}

\begin{lemma}\label{lemma:copyAll:AllNeighborsAreThere}
	Each process ``owns'' one distinct cell-list cell. The $copy^{ALL}_{\tilde{g}}$ function copies the particles from all neighbor cells/processes. 
	After the copy procedure, the particle storage compartments of all processes contain the particles of the corresponding cell and copies of the particles from all neighboring cells. Therefore, after each process has executed the copy function, all interaction partners of all particles in the center cell are in process-local memory.
\end{lemma}

\begin{lemma}\label{lemma:distPlacesParticlesOnlyInsideStorages}
	The functions $dist_{\left(\tilde{g},\mathcal{G}^1\right)}^{ALL}$ redistributes particles from the center particle storage compartment to only the other storage compartments on the same process. Therefore, it does not try to place them into non-existing storage compartments. Hence,
	\begin{multline}
		\forall w\in \{1,\ldots,N_{\text{cell}}\}\forall p^1_j \in \left<\left<\mathcal{P}^1\right>_w\right>_{\tfrac{3^d+1}{2}}: 
		p^2_j:=  \left<\left<\left< 
		step_{\mathcal{G}^1}^{ALL}\left(copy_{\tilde{g}}^{ALL}(\mathcal{P}^1)\right)
		\right>_w\right>_{\tfrac{3^d+1}{2}}\right>_j\\
		\longrightarrow\ 
		\alpha =  {}^{\ovun 3} \iota^{-1}\left(\left\lfloor \tfrac{1}{r_{c}}(\underline{x}_j^2 -\underline{D}_{\min})\right\rfloor - {}^{\ovun I} \iota ( w) +\ovun 3 \right)\in \{1,\ldots,3^d\}.
	\end{multline}
	
\end{lemma}

\begin{lemma}\label{lemma:distPlacesParticlesOnlyInsideDomain}
	The function $dist_{\left(\tilde{g},\mathcal{G}^1\right)}^{ALL}$ places particles only into storage compartments that represent cells inside the computational domain. Hence, processes at the domain border do not have particles in their ``outer'' storage compartments.
	Be
	\begin{equation}
		\underline{w}=\left(w_1,\ldots,w_d\right)^{\mathbf{T}}:=  {}^{\ovun I} \iota ( w)
	\end{equation}
	\begin{equation}
		\underline{\alpha}=\left(\alpha_1,\ldots,\alpha_d\right)^{\mathbf{T}}:= \left\lfloor \tfrac{1}{r_{c}}(\underline{x}_j^2 -\underline{D}_{\min})\right\rfloor - \underline w +\ovun 3,
	\end{equation}
	
	then
	\begin{multline}
		\forall w\in \{1,\ldots,N_{\mathrm{cell}}\} \quad \forall p^2_j:=  \left<\left<\left< 
		step_{\mathcal{G}^1}^{ALL}\left(copy_{\tilde{g}}^{ALL}(\mathcal{P}^1)\right)
		\right>_w\right>_{\tfrac{3^d+1}{2}}\right>_j:\\ 
		(k\in \{1,\ldots,d\} \land w_k=1) \ \rightarrow \ \alpha_k\in \{2,3\}\
		\land (k\in \{1,\ldots,d\} \land w_k=I_k)\ \rightarrow \ \alpha_k\in \{1,2\}.
	\end{multline}
\end{lemma}

\begin{lemma}\label{lemma:collectDistRestoreInitialDistributionCondition}
	\sloppy
	$\overline{\mathcal{P}}^1 :=step_{\mathcal{G}^1}^{ALL}\left(copy_{\tilde{g}}^{ALL}(\mathcal{P}^1)\right)$ does not necessarily fulfill the condition in Eq.~(\ref{eq:initialCell:ParticleDistribution}) to iterate $step_{\mathcal{G}^2}^{ALL}\left(copy_{\tilde{g}}^{ALL}\left(\overline{\mathcal{P}}^1\right)\right)$. 
	To avoid that $interact_g$ misses interactions due to incomplete neighborhoods, the particles in $\overline{\mathcal{P}}^1$ therefore need to be redistributed such that Eq.~(\ref{eq:initialCell:ParticleDistribution}) is fulfilled. 
	The decomposition of the initial permuted particle tuple is stored in the center storage compartment of the processes. Hence, Eq.~(\ref{eq:initialCell:ParticleDistribution}) can be rewritten for all state transition steps as 
	\begin{equation}\label{eq:particleCellPlacementCondition}
		\forall p^t_j\in 
		\underset{v=1}{\overset{N_{\mathrm{cell}}}{\fullmoon }}\ \left<\left< \mathcal{P}^t\right>_v\right> _{\tfrac{3^d+1}{2}}: 
		p^t_j \in \left<\left< \mathcal{P}^t\right>_w\right> _{\tfrac{3^d+1}{2}},
	\end{equation}
	where 
	\begin{equation}\label{eq:particleCellPlacementCondition:w}
		w= {}^{\ovun{I}}\iota^{-1}\left(\left \lfloor \frac{1}{r_c} ( \underline{x}^t_j-\underline{D}_{\min})\right \rfloor + \ovun{1} \right).
	\end{equation}
	This is achieved by the two functions $dist_{\left(\tilde{g} ,\mathcal{G}^1\right)}^{ALL}$ and $collect_{\tilde{g}}^{ALL}$.
\end{lemma}

\subsection{Proof of correctness and equivalence with sequential algorithm}
\begin{theorem}
	On the particles stored in the center storage compartments of the processes, and under the assumptions stated at the beginning of this section, the presented parallelization scheme for particle methods on distributed-memory computers computes the same results, except for particle ordering, as the underlying sequential particle method. Hence,
	\begin{equation}
		S\left(\left[ g^{1} ,\mathbf{p}^{1}\right]\right) =\left[\left< \mathcal{G}^{T}\right> _{1} \ ,\sigma ^{*}\left(\underset{w=1}{\overset{N_{\text{cell}}}{\fullmoon }}\left< \left< \mathcal{P}^{T}\right> _{w}\right> _{\tfrac{3^{d} +1}{2}}\right)\right]
	\end{equation}
	
	where 
	\begin{equation}
		\left[\mathcal{G}^{T} ,\mathcal{P}^{T}\right] =\tilde{S}\left(\left[\mathcal{G}^{1} ,\mathcal{P}^{1}\right]\right).
	\end{equation}
	
\end{theorem}

We prove that each sequential state transition step $s$ (eq.~\ref{eq:defPM:ss}) is equivalent to each parallel state transition step $\tilde{s}$ (eq.~\ref{eq:distributed:stateTransitionStep}) and that both algorithms terminate after the same number of state transitions. The sequential state transition step was defined as follows:
\begin{equation}
	\tag{eq. \ref{eq:defPM:ss}}
	s\left([ g,\mathbf{p}]\right)  :=\left[\overset{\circ }{e}(\overline{g}) ,\overline{\mathbf{p}}\right]\quad \text{with }[\overline{g} ,\overline{\mathbf{p}}] =\evoII\left( g,\intIII \left([ g,\mathbf{p}]\right)\right).
\end{equation}
Under the  condition the interact function (eq.~\ref{eq:condition:interactionPreviousInteractionIndependence}) and the neighborhood function (eq.~\ref{eq:condition:neighborhoodPreviousInteractionIndependence})  are independent of previous interactions, we can rewrite the third interact subfunction $\intIII$ for pull interaction particle methods (eq.~\ref{eq:condition:pullInteraction}) to \cite{Bamme2021}:
\begin{equation}
	\intIII \left([ g,\mathbf{p}]\right)=
	\left(\begin{array}{ c }
		p_{1} *_{_{1} i_{g}}\langle \mathbf{p}\rangle _{u( g,\mathbf{p} ,1)}\\
		\vdots \\
		\ p_{| \mathbf{p}| } *_{_{1} i_{g}}\langle \mathbf{p}\rangle _{u\left( g,\mathbf{p} ,| \mathbf{p}| \right)}
	\end{array}\right)^{\!\!\top}.
\end{equation}
We also know \cite{Bamme2021} that if the evolve function does not change the global variable (eq.~\ref{eq:condition:evolutionDoesntChangeGlobalVariable}), we can write the state transition step as:
\begin{equation}
	s\left([ g,\mathbf{p}]\right)  :=\left[\overset{\circ }{e}(g) ,\
	{}_{2}\evoII\left(\left(
	\begin{array}{ c }
		p_{1} *_{_{1} i_{g}}\langle \mathbf{p}\rangle _{u( g,\mathbf{p} ,1)}\\
		\vdots \\
		\ p_{| \mathbf{p}| } *_{_{1} i_{g}}\langle \mathbf{p}\rangle _{u\left( g,\mathbf{p} ,| \mathbf{p}| \right)}
	\end{array}\right)^{\!\!\top}
	\right)
	\right].
\end{equation}
We use proof by induction to prove equivalence for the global variable and the particles separately. We start with the global variable.  
Following the construction of $\mathcal{G}^1$, we directly get 
\begin{equation}
	\left(\begin{array}{ c }
		g^2\\
		\vdots \\
		g^2
	\end{array}\right)^{\!\!\top}
	=
	\left(\begin{array}{ c }
		\overset{\circ }{e}\left(g ^{1}\right)\\
		\vdots \\
		\overset{\circ }{e}\left(g^1\right)
	\end{array}\right)^{\!\!\top}
	=
	\left(\begin{array}{ c }
		\overset{\circ }{e}\left(\left< \mathcal{G}^1\right> _{1}\right)\\
		\vdots \\
		\overset{\circ }{e}\left(\left< \mathcal{G}^1\right> _{N_{\text{cell}}}\right)
	\end{array}\right)^{\!\!\top}
	=
	\left(\begin{array}{ c }
		\left< \mathcal{G}^2\right> _{1}\\
		\vdots \\
		\left< \mathcal{G}^2\right> _{N_{\text{cell}}}
	\end{array}\right)^{\!\!\top}
	=
	\mathcal{G}^2 .
\end{equation}
This is the base case for the proof by induction. For the induction step, we start from
\begin{equation}
	\left(\begin{array}{ c }
		g^t\\
		\vdots \\
		g^t
	\end{array}\right)^{\!\!\top}
	=
	\mathcal{G}^t
\end{equation}
and get
\begin{equation}
	\left(\begin{array}{ c }
		g^{t+1}\\
		\vdots \\
		g^{t+1}
	\end{array}\right)^{\!\!\top}
	=
	\left(\begin{array}{ c }
		\overset{\circ }{e}\left(g ^{t}\right)\\
		\vdots \\
		\overset{\circ }{e}\left(g^t\right)
	\end{array}\right)^{\!\!\top}
	=
	\left(\begin{array}{ c }
		\overset{\circ }{e}\left(\left< \mathcal{G}^t\right> _{1}\right)\\
		\vdots \\
		\overset{\circ }{e}\left(\left< \mathcal{G}^t\right> _{N_{\text{cell}}}\right)
	\end{array}\right)^{\!\!\top}
	=
	\left(\begin{array}{ c }
		\left< \mathcal{G}^{t+1}\right> _{1}\\
		\vdots \\
		\left< \mathcal{G}^{t+1}\right> _{N_{\text{cell}}}
	\end{array}\right)^{\!\!\top}
	=
	\mathcal{G}^{t+1}.
\end{equation}
This completes the part for the global variable. 

We now prove that the sequential state transition function $S$ and the distributed memory state transition function $\tilde{S}$ stop after the same number of states.
For the state transition $S$, the stop function $f$ only depends on $g^t$. For $\tilde{S}$, $f$ depends only on $\left<\mathcal G^t\right>_1$. From this and $g^t=\left<\mathcal G^t\right>_1$ follows
\begin{equation}
	\underline{{}_1 S\left(\left[ g^{1} ,\mathbf{p}^{1}\right]\right) =g^T = \left< \mathcal{G}^{T}\right> _{1} =\left< {}_1 \tilde S\left(\left[ \mathcal G^{1} ,\mathcal P^{1}\right]\right)\right> _{1}}
\end{equation}

\begin{equation}
	{}_2 s\left(\left[ g^{1} ,\mathbf{p}^{1}\right]\right)
	=
	{}_{2}\evoII\left(\left(
	\begin{array}{ c }
		p^1_{1} *_{_{1} i_{g^1}}\langle \mathbf{p}^1\rangle _{u( g^1,\mathbf{p}^1 ,1)}\\
		\vdots \\
		\ p^1_{| \mathbf{p}^1| } *_{_{1} i_{g^1}}\langle \mathbf{p}^1\rangle _{u\left( g^1,\mathbf{p}^1 ,| \mathbf{p}^1| \right)}
	\end{array}\right)^{\!\!\top}
	\right)
\end{equation}
The interact function is order-independent (eq.~\ref{eq:constraint:orderindependence}), and the permutation $\pi(\mathbf p^1)$ (eq.~\ref{eq:initialParticlesStorages:permutation}) of the initial particle tuple sorts the particles into cells. The neighborhood function is restricted to not using indices (eq.~\ref{eq:constraint:neighborhood}). Hence, for a permuted particle tuple, the neighborhood function returns the same result as for the un-permuted particle tuple, with the same permutation also applied to the result. This leads to     
\begin{equation}
	{}_2 s\left(\left[ g^{1} ,\mathbf{p}^{1}\right]\right)
	=
	{}_{2}\evoII\left(
	\pi^{-1}\left(
	\left(
	\begin{array}{ c }
		p^1_{\pi(1)} *_{_{1} i_{g^1}}\langle \pi(\mathbf{p}^1)\rangle _{u( g^1,\pi(\mathbf{p}^1) ,\pi(1))}\\
		\vdots \\
		\ p^1_{\pi(| \mathbf{p}^1|) } *_{_{1} i_{g^1}}\langle \pi(\mathbf{p}^1)\rangle _{u\left( g^1,\pi(\mathbf{p}^1) ,\pi(| \mathbf{p}^1|) \right)}
	\end{array}
	\right)^{\!\!\top}
	\right)
	\right).
\end{equation}
From the construction of the initial particle storages of all processes $\mathcal{P}^1$ (eqs.~\ref{eq:initialCell:ParticleDistribution} to \ref{eq:initialParticlesStorages:AllProcessors}), as well as Lemmata~\ref{lemma:copyAll:NoReadConflicts} and \ref{lemma:copyAll:AllNeighborsAreThere}, we derive 
\begin{equation}\label{eq:proofDistributedMemory:usedCopyAll}
	{}_{2} s\left(\left[ g^{1} ,\mathbf{p}^{1}\right]\right) 
	={}_{2} \evoII\left( \pi ^{-1}\left(\underset{w=1}{\overset{N_{\text{cell}}}{\fullmoon }}\left(
	\begin{array}{ c }
		^{w} p_{1}^{1} *_{_{1} i_{g^{1}}} {}^{w}\mathbf{q}^{1}{}_{u\left( g^{1} ,{}^{w}\mathbf{q}^{1} , {}^{w}z+1\right)}\\
		\vdots \\
		\ ^{w} p_{{}^{w}n}^{1} *_{_{1} i_{g^{1}}} {}^{w}\mathbf{q}^{1}{}_{u\left( g^{1} ,{}^{w}\mathbf{q}^{1} , {}^{w}z+{}^{w}n\right)}
	\end{array}\right)^{\!\!\top}\right)\right) ,
\end{equation}
where the tuple of all particles in the storage of the $w$-th process is 
\begin{equation}\label{eq:proofDistributedMemory:usedCopyAll:w}
	{}^{w}\mathbf{q}^{1} := \underset{l=1}{\overset{3^{d}}{\fullmoon }}\left< \left<copy_{\tilde{g}}^{ALL}(\mathcal{P}^1)\right>_w \right>_{l},
\end{equation}
the number of particles in ${}^{w}\mathbf{q}^{1}$ in front of the particles of the central storage compartment is
\begin{equation}
	{}^{w}z:=\sum _{l=1}^{\tfrac{3^{d} +1}{2} -1}\left| \left< \left<copy_{\tilde{g}}^{ALL}(\mathcal{P}^1)\right>_w\right> _{l}\right|,
\end{equation}
the number of particles in the central storage compartment is
\begin{equation}
	{}^{w}n=\left| \left< \left<copy_{\tilde{g}}^{ALL}(\mathcal{P}^1)\right>_w\right> _{\tfrac{3^{d} +1}{2}}\right|,
\end{equation}
and the particles in the central storage compartment are
\begin{equation}
	\left({}^{w} p_{1}^{1} ,\ldots,{}^{w} p_{{}^wn}^{1}\right) =\left< \left<copy_{\tilde{g}}^{ALL}(\mathcal{P}^1)\right>_w\right> _{\tfrac{3^{d} +1}{2}}.
\end{equation}
The interactions of the particles with their neighbors in eq.~\ref{eq:proofDistributedMemory:usedCopyAll} resembles the $interaction_{g}$ function (eq.~\ref{eq:stepAll:interaction}))
\begin{equation}
	{}_{2} s\left(\left[ g^{1} ,\mathbf{p}^{1}\right]\right) 
	={}_{2} \evoII\left( \pi ^{-1}\left(
	\underset{w=1}{\overset{N_{\text{cell}}}{\fullmoon }}
	interaction_{g^1}\left(\left<copy_{\tilde{g}}^{ALL}(\mathcal{P}^1)\right>_w\right)
	\right)\right).
\end{equation}
The evolution function cannot change the global variable (condition in eq.~\ref{eq:condition:evolutionDoesntChangeGlobalVariable}). Hence, the second evolution subfunction $\evoII$ can not change the global variable either and is, therefore, independent of the ordering of the particles. Since the evolve function $e$ can create or destroy particles, the permutation $\pi^{-1}$ can not rearrange the result of $\evoII$ to the sequential state transition result. But the calculation on the particles is identical. Hence, there exists a permutation $\tilde{\pi}^{-1}$ that rearranges the result such that
\begin{equation}
	{}_{2} s\left(\left[ g^{1} ,\mathbf{p}^{1}\right]\right) 
	= \tilde\pi ^{-1}\left( {}_{2}\evoII\left( 
	\underset{w=1}{\overset{N_{\text{cell}}}{\fullmoon }}
	interaction_{g^1}\left(\left<copy_{\tilde{g}}^{ALL}(\mathcal{P}^1)\right>_w\right)
	\right)\right).
\end{equation}
All processes calculate the $intercation_{g}$ function independently. Hence, the $\evoII$ function can also be executed on each process independently. This means that 
\begin{equation}
	{}_{2} s\left(\left[ g^{1} ,\mathbf{p}^{1}\right]\right) 
	= \tilde\pi ^{-1}\left(  
	\underset{w=1}{\overset{N_{\text{cell}}}{\fullmoon }}\
	{}_{2}\evoII\left(interaction_{g^1}\left(\left<copy_{\tilde{g}}^{ALL}(\mathcal{P}^1)\right>_w\right)
	\right)\right).
\end{equation}
The combination of ${}_{2}\evoII$ and $interaction_{g}$ is the same as the $step_{(\tilde{g},g)}$ function (eq.~\ref{eq:stepAll:step}) at the center storage compartment
\begin{equation}
	{}_{2} s\left(\left[ g^{1} ,\mathbf{p}^{1}\right]\right) 
	= \tilde\pi ^{-1}\left(  
	\underset{w=1}{\overset{N_{\text{cell}}}{\fullmoon }}\
	\left< step_{(\tilde{g},g^1)}\left(\left<copy_{\tilde{g}}^{ALL}(\mathcal{P}^1)\right>_w\right)\right> _{\tfrac{3^d+1}{2}}
	\right).
\end{equation}
The function $step_{\mathcal{G}^1}^{ALL}$ (eq. \ref{eq:stepAll:stepAll}) calculates the step for all processes, leading to
\begin{equation}
	{}_{2} s\left(\left[ g^{1} ,\mathbf{p}^{1}\right]\right) 
	= \tilde\pi ^{-1}\left(  
	\underset{w=1}{\overset{N_{\text{cell}}}{\fullmoon }}\
	\left<\left< step_{\mathcal{G}^1}^{ALL}\left(copy_{\tilde{g}}^{ALL}(\mathcal{P}^1)\right)\right>_w\right> _{\tfrac{3^d+1}{2}}
	\right).
\end{equation}
$\overline{\mathcal{P}}^1 :=step_{\mathcal{G}^1}^{ALL}\left(copy_{\tilde{g}}^{ALL}(\mathcal{P}^1)\right)$ does not necessarily fulfill the condition in eq.~\ref{eq:initialCell:ParticleDistribution}) to iterate  $step_{\mathcal{G}^2}^{ALL}\left(copy_{\tilde{g}}^{ALL}\left(\overline{\mathcal{P}}^1\right)\right)$. 
For $interact_g$ to not miss interactions due to incomplete neighborhoods, the particles in $\overline{\mathcal{P}}^1$ need to be redistributed such that the condition in eq.~\ref{eq:initialCell:ParticleDistribution} is fulfilled. This is achieved by the functions $dist_{\left(\tilde{g} ,\mathcal{G}^1\right)}^{ALL}$ (eq.~\ref{eq:distAll:distAll}) and $collect_{\tilde{g}}^{ALL}$ (eq.~\ref{eq:collectAll:collectAll}), as proven in the Lemmata \ref{lemma:distPlacesParticlesOnlyInsideStorages},  \ref{lemma:distPlacesParticlesOnlyInsideDomain}, and  \ref{lemma:collectDistRestoreInitialDistributionCondition}.
Hence, 
\begin{multline}
	{}_{2} s\left(\left[ g^{1} ,\mathbf{p}^{1}\right]\right) \\
	= {\tilde{\pi}} '^{-1}  \left(  
	\underset{w=1}{\overset{N_{\text{cell}}}{\fullmoon }}\
	\left<\left< collect_{\tilde{g}}^{ALL}\left( dist_{\left(\tilde{g} ,\mathcal{G}^1\right)}^{ALL}\left( step_{\mathcal{G}^1}^{ALL}\left( copy_{\tilde{g}}^{ALL} (\mathcal{P}^1 )\right)\right)\right)\right>_w\right> _{\tfrac{3^d+1}{2}}
	\right),
\end{multline}
where  ${\tilde{\pi}} '^{-1}$ is a new permutation. We insert the definition of ${}_{2}\tilde{s}_{\tilde{g}}$ to get
\begin{equation}
	{}_{2} s\left(\left[ g^{1} ,\mathbf{p}^{1}\right]\right) 
	= {\tilde{\pi}} '^{-1}  \left(  
	\underset{w=1}{\overset{N_{\text{cell}}}{\fullmoon }}\
	\left<\left< \underbrace{{}_{2}\tilde{s}_{\tilde{g}}\left([\mathcal{G}^1 ,\mathcal{P}^1]\right)}_{=:\mathcal{P}^2} \right>_w\right> _{\tfrac{3^d+1}{2}}
	\right),
\end{equation}
\begin{equation}
	\left[ g^{2} ,\mathbf{p}^{2}\right]
	= {\tilde{\pi}} '^{-1}  \left(  
	\underset{w=1}{\overset{N_{\text{cell}}}{\fullmoon }}\
	\left<\left< \mathcal{P}^2 \right>_w\right> _{\tfrac{3^d+1}{2}}
	\right).
\end{equation}
This is the base case for the proof by induction. For the induction step, we start from
\begin{equation}
	\left[ g^{t} ,\mathbf{p}^{t}\right]
	= {\tilde{\pi}} ''^{-1}  \left(  
	\underset{w=1}{\overset{N_{\text{cell}}}{\fullmoon }}\
	\left<\left< \mathcal{P}^t \right>_w\right> _{\tfrac{3^d+1}{2}}
	\right).
\end{equation}
We can assume that $\mathcal{P}^t$ fulfills the same conditions as $\mathcal{P}^2$ and, hence, also as $\mathcal{P}^1$,
especially the condition in eq.~\ref{eq:initialCell:ParticleDistribution} (or eqs.~\ref{eq:proofDistributedMemory:usedCopyAll}, \ref{eq:proofDistributedMemory:usedCopyAll:w}). Then, we can define a new particle method where $\left[ g^{t},\mathbf{p}^{t}\right]$ is the instance and $\mathcal{P}^t$ its corresponding cell-list-based distribution onto processes. Using Lemmata~\ref{lemma:distPlacesParticlesOnlyInsideStorages},  \ref{lemma:distPlacesParticlesOnlyInsideDomain}, and  \ref{lemma:collectDistRestoreInitialDistributionCondition} we derive that
\begin{equation}
	{}_{2} s\left(\left[ g^{t} ,\mathbf{p}^{t}\right]\right) 
	= {\tilde{\pi}} '''^{-1}  \left(  
	\underset{w=1}{\overset{N_{\text{cell}}}{\fullmoon }}\
	\left<\left< \underbrace{{}_{2}\tilde{s}_{\tilde{g}}\left([\mathcal{G}^t ,\mathcal{P}^t]\right)}_{=:\mathcal{P}^{t+1}} \right>_w\right> _{\tfrac{3^d+1}{2}}
	\right),
\end{equation}
hence,
\begin{equation}
	\left[ g^{t+1} ,\mathbf{p}^{t+1}\right]
	= {\tilde{\pi}} '''^{-1}  \left(  
	\underset{w=1}{\overset{N_{\text{cell}}}{\fullmoon }}\
	\left<\left< \mathcal{P}^{t+1} \right>_w\right> _{\tfrac{3^d+1}{2}}
	\right).
\end{equation}
We do this now for all $t$ until $f{g^t}=\top$. Together with the derivation of the proof for the global variable, this leads to
\begin{equation}
	S\left(\left[ g^{1} ,\mathbf{p}^{1}\right]\right) =\left[\left< \mathcal{G}^{T}\right> _{1} \ ,\sigma ^{*}\left(\underset{w=1}{\overset{N_{\text{cell}}}{\fullmoon }}\left< \left< \mathcal{P}^{T}\right> _{w}\right> _{\tfrac{3^{d} +1}{2}}\right)\right] ,
\end{equation}
where $ \left[\mathcal{G}^{T} ,\mathcal{P}^{T}\right] =\tilde{S}\left(\left[\mathcal{G}^{1} ,\mathcal{P}^{1}\right]\right) $.

Hence, the present parallelization scheme produces the same result up to a different ordering of the particles. The particles are permuted by an unknown permutation $\sigma^*$. This proves that the distributed-memory parallelization scheme is correct for order-independent particle methods. 

\section{Bounds on Time Complexity and Parallel Scalability}\label{sec:distributedparallel:timeComplexety}
An algorithm's time complexity describes the runtime required by a machine to execute that algorithm. It depends on the input size of the algorithm. For a particle method, the input size is the length of the initial tuple $\mathbf p^1$. We assume that constants bind the sizes of the global variable and each particle.
We further assume that the algorithm terminates in a finite time. Hence, an upper bound exists for all functions' time complexities. Following \cite{Bamme2021}, an upper bound on the time complexity of the interact function $\tau _{ i( g,p',p'')}$ is
\begin{equation}
	\forall g \in \{g^1,\ldots,g^T\}, p',p'' \in \underset{w=1}{\overset{T}{\fullmoon }} \mathbf{p}^w\ 
	\exists \tilde g \in \{g^1,\ldots,g^T\}, \tilde p',\tilde p'' \in \underset{w=1}{\overset{T}{\fullmoon }} \mathbf{p}^w:\
	\tau _{ i( g ,p',p'')} \leq \tau_{ i(\tilde{g} ,\tilde{p} ',\tilde{p} '')}=:\tau _{i}.
\end{equation}
Similarly, an upper bound on the time complexity of the evolve function $\tau_{ e( g,p)}$ is
\begin{equation}
	\forall g \in \{g^1,\ldots,g^T\}, p \in \underset{w=1}{\overset{T}{\fullmoon }} \mathbf{p}^w\ 
	\exists \tilde g \in \{g^1,\ldots,g^T\}, \tilde p \in \underset{w=1}{\overset{T}{\fullmoon }} \mathbf{p}^w:\
	\tau_{ e( g ,p)} \leq \tau_{ e(\tilde{g} ,\tilde{p})}=:\tau _{e}.
\end{equation}
An upper bound on the time complexity of the evolve function of the global variable $\tau_{\overset{\circ }{e}( g)}$ is
\begin{equation}
	\forall g \in \{g^1,\ldots,g^T\}\ 
	\exists \tilde g \in \{g^1,\ldots,g^T\}:\quad
	\tau_{\overset{\circ }{e}( g)} \leq \tau_{\overset{\circ }{e}(\tilde{g})}=:\tau _{\overset{\circ }{e}}.
\end{equation}
An upper bound on the time complexity of the stopping condition function $	\tau _{ f( g)}$ is
\begin{equation}
	\forall g \in \{g^1,\ldots,g^T\}\ 
	\exists \tilde g \in \{g^1,\ldots,g^T\}:\quad
	\tau _{ f( g)} \leq \tau_{f(\tilde{g})}=:\tau _{f}.
\end{equation}
An upper bound on the time complexity of the neighborhood function $\tau_{  u( [g,\mathbf{p} ],j)}$ is
\begin{multline}
	\forall g \in \{g^1,\ldots,g^T\}, \mathbf{p} \in \{\mathbf{p}^1,\ldots, \mathbf{p}^T\}, j \in \{1,\ldots,|\mathbf{p}|\}\\
	\exists \tilde g \in \{g^1,\ldots,g^T\},\tilde{ \mathbf{p}} \in \{\mathbf{p}^1,\ldots, \mathbf{p}^T\}, \tilde j \in \{1,\ldots,|\tilde{ \mathbf{p}}|\}:\\
	\tau_{  u( [g,\mathbf{p} ],j)} \leq \tau_{ u( [\tilde{g} ,\tilde{\mathbf{p}} ],\tilde{j})} =:\tau _{u}.
\end{multline}
An upper bound on the size of the neighborhood $| u( [g,\mathbf{p} ],j)|$ is
\begin{multline}
	\forall g \in \{g^1,\ldots,g^T\}, \mathbf{p} \in \{\mathbf{p}^1,\ldots, \mathbf{p}^T\}, j \in \{1,\ldots,|\mathbf{p}|\}\\
	\exists \tilde g \in \{g^1,\ldots,g^T\},\tilde{ \mathbf{p}} \in \{\mathbf{p}^1,\ldots, \mathbf{p}^T\}, \tilde j \in \{1,\ldots,|\tilde{ \mathbf{p}}|\}:\\
	| u( [g,\mathbf{p} ],j)| \leq | u( [\tilde{g} ,\tilde{\mathbf{p}} ],\tilde{j})| =:\varsigma_{u}.
\end{multline}
The time complexity of calculating ${}^{\ovun{I}}\iota$ and ${}^{\ovun{I}}\iota^{-1}$ is in $\mathcal{O}(d)$. 
Therefore, the time complexity of the index transformation functions is bound by $C d$, where $C$ is a constant.
Hence, the time complexity of $\beta$,$\tau_{\beta(t,l)}$ is bound by 
\begin{equation}
	\forall\ \ovun{I} \in \mathbb{N}^d_{>0}, \forall t\in \{1,\ldots,N_{\text{cell}}\}, \,\, l \in \{1,\ldots,3^d\}\ \exists C_{\beta} \in \mathbb{ R}:\ \tau_{\beta(t,l)}\leq C_{\beta} d.
\end{equation}
The time complexity of $\gamma$,$\tau_{\gamma(t,l)}$, is bound by
\begin{equation}
	\forall\ \ovun{I} \in \mathbb{N}^d_{>0}, \forall k \in \{1,\ldots,3^d\},\,\, j \in \{1,\ldots,{}^{k}N^*_{\text{cell}}\} \ \exists C_{\gamma} \in \mathbb{ R}:\ \tau_{\gamma(t,l)}\leq C_{\gamma} d.
\end{equation}
The time complexity of computing $\alpha$ is bound by
\begin{equation}
	\forall\ \underline{D}_{\min},\underline{D}_{\max}\in \mathbb{R}^d, \underline{x}\in \left[\underline{D}_{\min},\underline{D}_{\max}\right],\,\, \forall j\in \{1,\ldots,N_{\text{cell}}\}\ \exists C_{\alpha} \in \mathbb{R}:\
	\tau_{{}^{\ovun 3} \iota^{-1}\left(\left\lfloor \tfrac{1}{r_{c}}(\underline{x} -\underline{D}_{\min})\right\rfloor - {}^{\ovun I} \iota ( j) +\ovun 3 \right)}\leq C_{\alpha} d.
\end{equation}
The time complexity for computing the index transformation in the $collect$ function is bound by
\begin{equation}
	\forall l \in \{1,\ldots,3^d\}\ \exists C_{c} \in \mathbb{R}:
	\tau_{{}^{\ovun 3} \iota^{-1}\left(\ovun 4 -{}^{\ovun 3} \iota ( l)\right)}\leq C_{c} d.
\end{equation}
We use these upper bounds to derive bounds on the time complexity of the present parallelization scheme on a sequential computer and on a distributed-memory parallel machine. In general, an upper bound on the time complexity of the state transition depends on the instance $[g^1,\mathbf{p}^1]$. For each instance, we can  bound the number of particles by
\begin{equation}
	\forall t \in \{1,\ldots,T\}\ \exists N^{\max}_{\mathbf p}\in \mathbb{N}: \left|\mathbf{p}^t\right| \leq N^{\max}_{\mathbf p}.
\end{equation}
The time complexity of the sequential state transition $\tau_S$ is then bound by:
\begin{equation}  \label{eq:timeCompletity:stdStateTransition}                     
	\tau_{ S\left(\left[ g^{1} ,\mathbf{p}^{1}\right]\right)}  
	\leq  T \left(N^{\max}_{\mathbf p}  \left( \varsigma_{u}\left(N^{\max}_{\mathbf p}\right)\ \tau _{i} +\tau _{u}\left(N^{\max}_{\mathbf p}\right) +\tau _{e}\right) +\tau _{f} +\tau _{\overset{_{\circ }}{e}}\right),
\end{equation}
where the neighborhood-related terms $\varsigma_{u}\left(N^{\max}_{\mathbf p}\right)$ and $\tau _{u}\left(N^{\max}_{\mathbf p}\right)$ potentially depend on $N^{\max}_{\mathbf p}$.
In the presented cell list-based parallelization scheme, we can further bind the neighborhood function by exploiting that the neighborhood calculation is done separately on each process. Hence, only the particles in that process are taken into account. The number of particles in one cell is bound by
\begin{equation}
	\forall t \in \{1,\ldots,T\},\,\, w\in\{1,\ldots,N_{\text{cell}}\}\ \exists n_{\max}\in\mathbb{N}:\
	\left|\left< \left< \mathcal{P}^t\right>_w\right>_{\tfrac{3^d+1}{2}}\right|\leq n_{\max}.
\end{equation}
Then, the number of all particles in one process is bound by 
\begin{equation}
	\forall t \in \{1,\ldots,T\},\,\, w\in\{1,\ldots,N_{\text{cell}}\}:\
	\left|\underset{l=1}{\overset{3^d}{\fullmoon }}\left< \left< \mathcal{P}^t\right>_w\right>_{l}\right| \leq 3^d n_{\max}.
\end{equation}
The number of particles is bound, and the neighborhood function checks if a distance is smaller than $r_c$ and verifies a function $\Omega(g,p_k,p_j)$. We assume
\begin{equation}
	\forall g \in \{g^1,\ldots,g^T\}, \mathbf{p} \in \{\mathbf{p}^1,\ldots, \mathbf{p}^T\}, j,k \in \{1,\ldots,|\mathbf{p}|\}\ \exists C_u\in\mathbb{R}:\
	\tau_{|\underline{x}_k-\underline{x}_j|\leq r_c}+\tau_{\Omega(g,p_k,p_j)}\leq d C_u
\end{equation} 
, then the time complexity and the size of the neighborhood function are bound by
\begin{equation}
	\tau_u \leq 3^d n_{\max} d C_u,\qquad \varsigma_{u} \leq 3^d n_{\max}.
\end{equation}

The time complexity of sequentially executing the distributed-memory parallel particle method $\tau _{\tilde{S}\left(\left[ g^{1},\mathbf{p}^{1}\right]\right)} (1)$ is determined by the time complexity of the functions $collect_{\tilde{g}}^{ALL}$, $dist_{\left(\tilde{g},\mathcal{G}\right)}^{ALL}$, $ step_{\mathcal{G}}^{ALL}$, $copy_{\tilde{g}}^{ALL}$, the evolve function of the global variable for each cell, and the stop function.
Then,
\begin{equation}
	\begin{array}{ r l l }
		\tau _{\tilde{S}\left(\left[ g^{1} ,\mathbf{p}^{1}\right]\right)} (1) & \leq T\biggl( & \tau _{f} +\underbrace{N_{\text{cell}}  \tau _{\overset{_{\circ }}{e}}}_{(global\ variable\ evolve)}\
		+\underbrace{\prod{}{_{k=1}^{3^{d}}} {}^{k}N_{\text{cell}}^{*}  \left( C_{\gamma }  d +3^{d}  (C_{\beta }  d+C_{c}  d+n_{\max} )\right)}_{(collect) }\\
		&  & +\underbrace{N_{\text{cell}}  n_{\max}  (C_{\alpha }  d+1)}_{(dist)}\
		+\underbrace{N_{\text{cell}}  n_{\max}  \left( \tau _{e} +3^{d}  n_{\max} C_u d  +3^{d}  n_{\max}  \tau _{i}\right)}_{(step)}\\
		&  & +\underbrace{\prod{}{_{k=1}^{3^{d}}}{}^{k} N_{\text{cell}}^{*}  \left( C_{\gamma }  d + 3^{d} (C_{\beta }  d+n_{\max} )\right)}_{(copy)}\biggr).
	\end{array}
\end{equation}
We assume that particles are evenly distributed (i.e., the number of particles in each cell is approximately the same) and that the density of particles remains constant when increasing the number of initial particles $\mathbf{p}^1$. Therefore, the domain has to increase.  
Hence, only the number of cells $N_{\text{cell}}$ increases, and $n_{\max}$ stays approximately the same. 
Under the assumption that all functions are then bound by constants, since they do not depend on $N_{\text{cell}}$, we can simplify
\begin{equation} \label{eq:timeComplexity:simplDistributed:raw}
	\tau _{\tilde{S}\left(\left[ g^{1} ,\mathbf{p}^{1}\right]\right)} (1)\leq T\biggl( C_{f} +N_{\text{cell}}  \left( C_{\overset{_{\circ }}{e}}  +C_{dist} +C_{step}\right)\
	+ \prod{}{_{k=1}^{3^{d}}}{}^{k} N_{\text{cell}}^{*} \left(C_{collect}+C_{copy}\right)\biggr).
\end{equation}
For a single processor we can further simplify by using  $\prod{}{_{k=1}^{3^{d}}}^{k} N_{\text{cell}}^{*} = N_{\text{cell}}$ to
\begin{equation} \label{eq:timeComplexity:simplDistributed:siglePROC}
	\tau _{\tilde{S}\left(\left[ g^{1} ,\mathbf{p}^{1}\right]\right)} (1)\leq T\left( C_{f} +N_{\text{cell}}  ( C_{\overset{_{\circ }}{e}} +C_{collect} +C_{dist} +C_{step} +C_{copy})\right).
\end{equation}
When parallelizing it, we need to consider that a process is the smallest computational unit.
The processes are distributed on a number of processors $n_{CPU}$. We define a ``processor'' (CPU) as running concurrently and operating on its own separate memory address space.
For convenience we assume $\frac{N_{cell}}{3^d}\in \mathbb{N}$.
Further, to avoid communications conflicts, the processes need to be distributed to the processors according to which checkerboard pattern they belong.
All processes of the $k$th-checkerboard pattern are the $\gamma ( k,j)$th-processes with $k$ fixed and $j\in\{1,...,{}^{k} N_{\text{cell}}^{*}\}$.
The processes on one processor must be either from one checkerboard pattern or the entire checkerboard pattern. 
We formulate the time complexity of the parallelized distributed-memory parallel particle method as 
\begin{multline} 
	\tau _{\tilde{S}\left(\left[ g^{1} ,\mathbf{p}^{1}\right]\right)} (n_{CPU} )\leq T\biggl( C_{f} +\Xi _{calc}( N_{cell} ,n_{CPU} ,d)( C_{\overset{_{\circ }}{e}} +C_{dist} +C_{step})
	\\
	+\Xi _{com}( N_{cell} ,n_{CPU} ,d)( C_{collect} +C_{copy})\biggr),
\end{multline}
where
\begin{equation}
	\Xi _{calc}( N_{cell} ,n_{CPU} ,d) :=\begin{cases}
		\left\lceil \frac{3^{d}}{n_{CPU}}\right\rceil \frac{N_{cell}}{3^{d}} & \text{if} \quad n_{CPU} \in \left\{1,...,3^{d}\right\}\\
		M_{2} & \text{if} \quad n_{CPU} \in \left\{3^{d} ,...,N_{cell}\right\}\\
		1 & \text{if} \quad n_{CPU}  >N_{cell},
	\end{cases}
\end{equation}
and
\begin{equation}
	\Xi _{com}( N_{cell} ,n_{CPU} ,d) :=\begin{cases}
		N_{cell} & \text{if} \quad n_{CPU} \in \left\{1,...,3^{d}\right\}\\
		n_{1} M_{1} +n_{2} M_{2} & \text{if} \quad n_{CPU} \in \left\{3^{d} ,...,N_{cell}\right\}\\
		3^{d} & \text{if} \quad n_{CPU}  >N_{cell}
	\end{cases},
\end{equation}
where the number of checkerboard patterns that have more processors assigned to them is
\begin{equation}
	n_{1} :=n_{CPU}\bmod 3^{d},
\end{equation}
the number of checkerboard patterns that have fewer processors assigned to them is
\begin{equation}
	n_{2} :=3^{d} -n_{1},
\end{equation}
the maximum number of processes per processor is
\begin{equation}
	M_{1} :=\left\lceil \tfrac{N_{cell}}{3^{d}\left\lceil \tfrac{n_{CPU}}{3^{d}}\right\rceil }\right\rceil,
\end{equation}
and for checkerboard patterns with fewer processors assigned to them, the maximum number of processes per processor is
\begin{equation}
	M_{2} :=\left\lceil \tfrac{N_{cell}}{3^{d}\left\lfloor \tfrac{n_{CPU}}{3^{d}}\right\rfloor }\right\rceil. 
\end{equation}
For $n_{CPU} \in \left\{1,...,3^{d}\right\}$, the number of processors did not reach the number of checkerboard patterns. In this case, each checkerboard pattern is completely on a processor to avoid communication conflicts. The limiting factor for the calculation ($\Xi_{calc}$) is then the maximum number of entire checkerboard patterns on one processor. This is $\left\lceil \frac{3^{d}}{n_{CPU}}\right\rceil \frac{N_{cell}}{3^{d}}$, where  $\left\lceil \frac{3^{d}}{n_{CPU}}\right\rceil$ is the maximum number of checkerboard pattern on one processor and  $ \frac{N_{cell}}{3^{d}}$ the number of processes in one checkerboard pattern. The communication ($\Xi_{com}$) is then sequential since a processor does the communication of each process sequentially, and the communication for each checkerboard pattern needs to be done sequentially.

For $n_{CPU} \in \left\{3^{d} ,...,N_{cell}\right\}$, a checkerboard pattern can be distributed on more the one processor. Therefore the limiting factor for the calculation ($\Xi_{calc}$) is the maximum number of processes per processor, which is $M_2$. In $M_2$, the term $\frac{N_{cell}}{3^{d}}$ is again the number of processes in one checkerboard pattern and $\left\lfloor \tfrac{n_{CPU}}{3^{d}}\right\rfloor$ is the minimum number of processors per checkerboard pattern. 
The limiting factor for the communication ($\Xi_{com}$) is more complex since the communication is carried out for each checkerboard pattern separately, one after the other. Hence, there is a number ($n_1$) of checkerboard patterns that have one processor more assigned ($\left\lceil \tfrac{n_{CPU}}{3^{d}}\right\rceil$) to them and a number ($n_2$) of checkerboard pattern with less ($\left\lfloor \tfrac{n_{CPU}}{3^{d}}\right\rfloor$). The processors for one checkerboard pattern can communicate in parallel but sequential for the checkerboard pattern. Hence to sum up the maximum number of communication per processor for all checkerboard patterns, we calculate $n_{1} M_{1} +n_{2} M_{2}$.

Each processor has exactly one process or no process for $n_{CPU}$ reaching or exceeding $N_{cell}$. The calculation ($\Xi_{calc}$) is saturated, and the limiting factor is $1$. The communication ($\Xi_{com}$) is also saturated, and the limiting factor is the number of checkerboard patterns $3^d$.

These upper bounds of the time complexities allow for investigations on the speed-ups we get with the cell list scheme on one processor, and the speed-ups regarding Amdahl's \cite{Amdahl1967}, and Gustafson's law \cite{Gustafson1988}.

First, the speed-up of the cell list scheme on one processor vs. the sequential state transition is
\begin{align}
	speedup_{cell}\left( N_{\mathbf{p}}^{\max}\right) &:=\frac{\tau _{S\left(\left[ g^{1} ,\mathbf{p}^{1}\right]\right)}}{\tau _{\tilde{S}\left(\left[ g^{1} ,\mathbf{p}^{1}\right]\right)}( 1)}
	\\[6pt]
	&\approx \frac{N{_{\mathbf{p}}^{\max}}^{2} C_u d +N_{\mathbf{p}}^{\max}\left( 3^{d} n_{\max} \tau _{i} +\tau _{e}\right) +\tau _{f} +\tau _{\overset{_{\circ }}{e}}}
	{N_{\mathbf{p}}^{\max}\left( 3^{d} n_{\max} C_u d +3^{d} n_{\max} \tau _{i} +\tau _{e} +\frac{\tau _{\overset{_{\circ }}{e}} +C_{collect} +C_{dist} +C_{copy}}{n_{\max}}\right) +\tau _{f}}
	\\[6pt]
	&\label{eq:complexity:celllist:speedup:order}\in \frac{\mathcal{O}\left( N{_{\mathbf{p}}^{\max}}^{2}\right)}
	{\mathcal{O}\left( N_{\mathbf{p}}^{\max}\right)} =\mathcal{O}\left( N_{\mathbf{p}}^{\max}\right).
\end{align}
Since particles can move, the neighborhood search function checks each pair of particles to see if they are neighbors. This has a complexity in $\mathcal{O}(n^2)$, where $n$ is the number of particles. Fast neighbor list algorithms like cell-lists \cite{Hockney:1988} reduce this to $\mathcal{O}(n)$ under the assumptions made here. We can confirm this by comparing the simplified time complexity of the cell-list-based scheme (eq.~\ref{eq:timeComplexity:simplDistributed:siglePROC}) with the time complexity of the sequential state transition (eq.~\ref{eq:timeCompletity:stdStateTransition}), where we eliminated the dependency of the time complexity of the neighborhood function on the particle number. Our cell-list-based scheme scales linearly with the number of particles ($N_{\text{cell}} n_{\max}\approx N^{\max}_{\mathbf p}$) and not quadratically as it would without the assumption of even and constant particle density. Hence, the speed-up is $\mathcal{O}\left( N_{\mathbf{p}}^{\max}\right)$ as derived in equation \ref{eq:complexity:celllist:speedup:order} and visualized in figure \ref{fig:complexity:speedups:celllist}.

Second, Amdahl's Law is the speed-up of the cell-list-based scheme by multiple processors, where the problem size is fixed for increasing processors $n_{\text{CPU}}$. 
\begin{align}
	speedup_{Amdal}( n_{CPU}) &=\frac{\tau _{\tilde{S}\left(\left[ g^{1} ,\mathbf{p}^{1}\right]\right)} (1)}{\tau _{\tilde{S}\left(\left[ g^{1} ,\mathbf{p}^{1}\right]\right)} (n_{CPU} )}
	\\
	&\approx \tfrac{ C_{f} +N_{cell} \ ( C_{\overset{_{\circ }}{e}} +C_{collect} +C_{dist} +C_{step} +C_{copy})}
	{ C_{f} +\Xi _{calc}( N_{cell} ,n_{CPU} ,d)( C_{\overset{_{\circ }}{e}} +C_{dist} +C_{step}) +\Xi _{com}( N_{cell} ,n_{CPU} ,d)( C_{collect} +C_{copy})}.
\end{align}
In this case, we can increase $n_{\text{CPU}}$ until we reach the number of cells $N_{\text{cell}}$. After that, there will be no further speed-up. But also, until then, we find a step-like behavior with increasing $n_{\text{CPU}}$ because cells cannot be split across CPUs as visualized in figure \ref{fig:complexity:speedups:amdal}.

Third, Gustafson's law is the speed-up of the cell-list-based scheme by multiple processors, where the ratio of problem size and CPU number is kept constant while increasing the number of processors. Thus $\tfrac{N_{\text{cell}}}{n_{\text{CPU}}}=const$.
We achieve this by setting $N_{\text{cell}}=n_{CPU}\cdot N^{'}_{\text{cell}}$, where $N^{'}_{\text{cell}}$ is constant.
For a perfectly fitting processor interconnect network topology, we predict a linear-like speed-up with steps as visualized in figure \ref{fig:complexity:speedups:gustafson}.
\begin{multline}
	speedup_{Gustafson}( n_{CPU}) =\frac{\tau _{\tilde{S}\left(\left[ g^{1} ,\mathbf{p}^{1}( n_{CPU})\right]\right)} (1)}{\tau _{\tilde{S}\left(\left[ g^{1} ,\mathbf{p}^{1}( n_{CPU})\right]\right)} (n_{CPU} )}
	\\
	\approx \tfrac{ C_{f} +n_{CPU} \cdot N_{cell} \ ( C_{\overset{_{\circ }}{e}} +C_{collect} +C_{dist} +C_{step} +C_{copy})}
	{ C_{f} +\Xi _{calc}( n_{CPU} \cdot N_{cell} ,n_{CPU} ,d)( C_{\overset{_{\circ }}{e}} +C_{dist} +C_{step}) +\Xi _{com}( n_{CPU} \cdot N_{cell} ,n_{CPU} ,d)( C_{collect} +C_{copy})}
\end{multline}
Overall the scheme behaves as expected for cell-list algorithms.
\newcommand\widthComp{56mm}
\newcommand\widthCompG{59mm}
\begin{figure}[H]
	\begin{subfigure}[t]{\widthComp}
		\centerline{\includegraphics[width=\widthCompG]{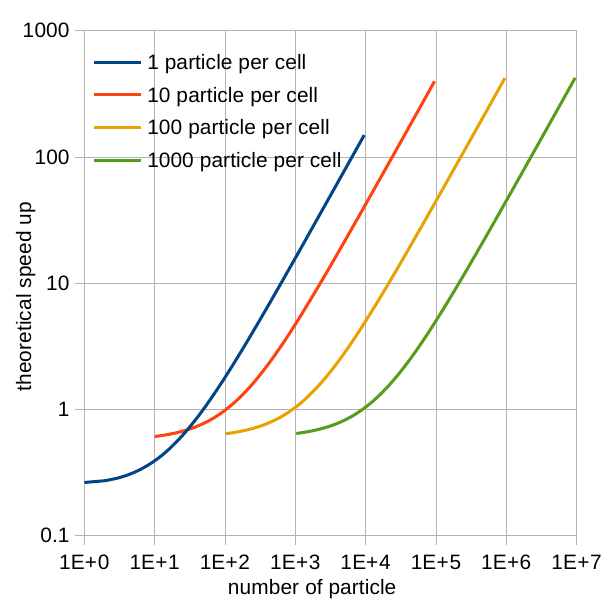}}
		\caption{Speed-up with a cell list algorithm. \label{fig:complexity:speedups:celllist}}
	\end{subfigure}\hspace{2mm}
	\begin{subfigure}[t]{\widthComp}
		\centerline{\includegraphics[width=\widthCompG]{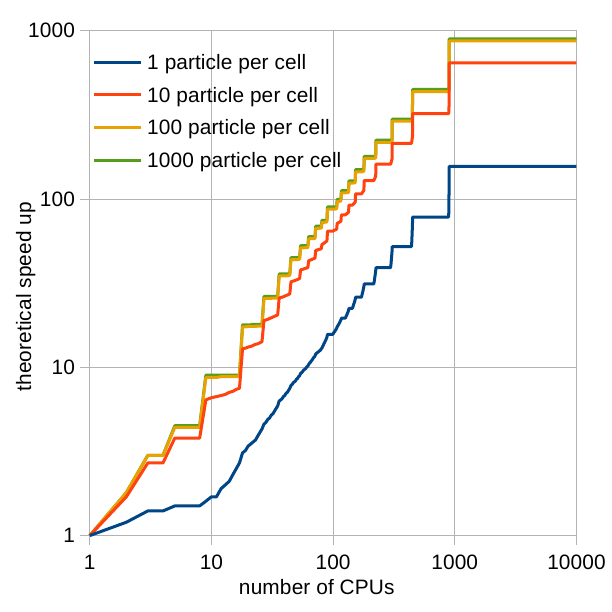}}
		\caption{Speed-up of Amdahl's law. \label{fig:complexity:speedups:amdal}}
	\end{subfigure}\hspace{2mm}
	\begin{subfigure}[t]{\widthComp}
		\centerline{\includegraphics[width=\widthCompG]{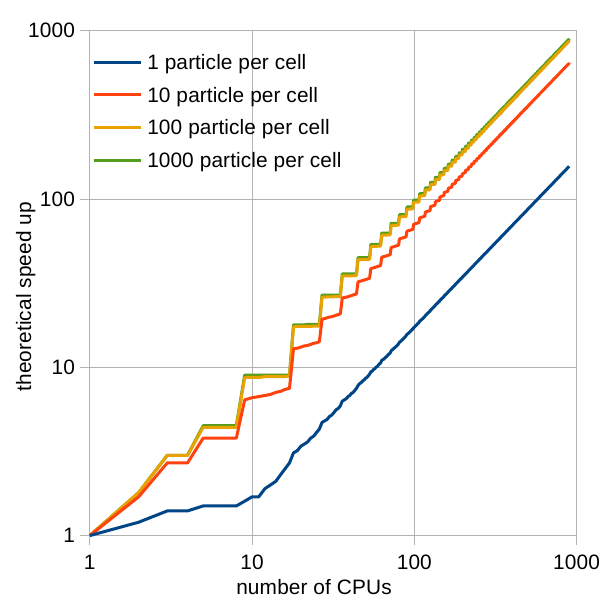}}
		\caption{Speed-up of Gustafson's law. \label{fig:complexity:speedups:gustafson}}
	\end{subfigure}
	\caption{Theoretical Speed-Ups. The constants are chosen to be $d=2$, $C_u=C_{\alpha}=C_{\beta}=C_{\gamma}=1$, $\tau_i=\tau_e=3$, and $\tau_f=\tau_{\overset{_{\circ }}{e}}=1$. For Amdahl's and Gustafson's law $N_{cell}=900$.  \label{fig:complexity:speedups}}
\end{figure}  

\section{Conclusions and Discussion} \label{sec:conclusion}

Particle methods encompass a wide range of computer simulation algorithms, such as Discrete-Element Methods (DEM) \cite{Walther:2009}, Molecular Dynamics (MD) \cite{Alder:1957}, Reproducing Kernel Particle Methods (RKPM) \cite{Liu:1995},  Particle Strength Exchange (PSE) \cite{Degond:1989a, Eldredge:2002}, Discretization-Corrected PSE (DC-PSE) \cite{Schrader:2010, Bourantas:2016}, and Smoothed Particle Hydrodynamics (SPH) \cite{Gingold:1977, Monaghan:2005}.
Although particle methods are widely used, mathematically defined, and practically parallelized in software frameworks such as the PPM Library \cite{Sbalzarini:2006b}, OpenFPM \cite{Incardona:2019}, POOMA \cite{Reynders:1996a}, or FDPS \cite{Iwasawa:2016}. However, little work has been done to formally investigate generic parallelization schemes for particle methods and prove their correctness and computational complexity.

Here, we took a first step in this direction by providing a shared-memory parallelization scheme for particle methods independent of specific applications. We proved the correctness of the presented cell-list-based parallelization scheme by showing equivalence to the sequential particle methods definition under certain assumptions, which we also defined here. Finally, we derived upper bounds on the time complexity of the proposed scheme executed on both sequential and parallel computers, and we discussed the parallel scalability. 
The presented parallelization scheme applies to an entire class of particle methods. We proved that it returns the same results as the original sequential definition of particle methods under certain assumptions. 

The presented parallelization scheme is not novel and similar to what is commonly implemented in software. Therefore, our analysis is of immediate practical relevance for general-purpose particle methods frameworks, even for critical calculation, as the proof guarantees correctness.
A limitation of the proposed scheme is that it assumes only pull interactions between the particles. This neglects the potential runtime benefits and versatility of symmetric interaction evaluations. However, pull interactions are suitable for more processor architectures, especially in a shared memory setting, which could allow for combining this distributed scheme with the shared memory scheme proven earlier \cite{Bamme2021}. Pull interactions also reduce communication since only particles in the center cell of a CPU are changed. They do not need to be communicated back, as would be the case for push or symmetric interactions, which also change copies of particles from other CPUs.
We also restricted the neighborhood function such that the cell-list strategy became applicable. This limits the expressiveness of the particle method but allows the efficient distribution of moving particles.  
Further, the constraint that particles are not allowed to leave the domain keeps domain handling simple. Otherwise, cells would dynamically need to be added or removed as necessary, resulting in a much more complex and dynamic mapping of cells to processors.
We also restricted particles to not moving further than the cutoff radius in a single state iteration or time step of the algorithm. Since the cutoff radius determines cell sizes, and individual cells cannot be split across multiple processes, this has the benefit that processes only need to communicate with their immediately adjacent neighbors. 
Additionally, we restricted the global variable only to be changed by the evolve function of the global variable and not by the evolve function of the particles. Therefore, no global operations are allowed where global variable changes would require global synchronization. Hence, the global variable change can be computed locally, still keeping it in sync even without communication. 
Finally, the time complexity of the checkerboard-like communication scheme has not an optimal prefactor, leaving many processes inactive. Nevertheless, it scales linearly with the number of processors.

Notwithstanding these limitations, with its proof of equivalence to the sequential particle methods definition, the present parallelization scheme stands in contrast with the primarily so-far algorithm-specific or experimentally tested parallelization schemes.
The rigorousness of our analysis paves the way for future research into the theory of parallel scientific simulation algorithms and the engineering of provably correct parallel software implementations.

Future theoretical work could optimize the presented scheme for computer architectures with parallel or synchronously clocked communication. Global operations could then also be included. Also, the network topology of the machine could be incorporated into the parallelization scheme. 
Furthermore, proofs for push and symmetric interaction schemes could be beneficial for specific use cases, as well as combining parallelization schemes for shared and distributed memory to better match the heterogeneous architecture of modern supercomputers. 
On the software engineering side, future work could leverage the presented parallelization scheme and proofs to design a new generation of theoretically founded software frameworks. They would potentially be more predictable, suitable for security- and safety-critical applications, and more maintainable and understandable as they are based on a common formal framework.

Overall, the presented proven parallelization scheme provides a way to parallelize a class of particle methods on distributed-memory systems with full knowledge of its validity and assumptions. We proved that it computes the same result as the underlying sequential particle method. Therefore, using it in a general framework for particle methods is suitable, even for critical computations, since the proof guarantees that the parallelization does not change the results.
We, therefore, hope that the present work will generate downstream investigation and studies in the theory of algorithms for scientific computing.

\section*{Acknowledgments}
We thank Pietro Incardona for discussions on distributed memory parallelization.

\appendix
\newpage
\section{Proofs of Lemmata}\label{sup:lemmata}
We prove the helping lemmata for the proof that the presented parallelization scheme is equivalent to the original particle method.

\begin{proof} \textbf{of lemma \ref{lemma:indexTransformIsBijectiv}}\\
	We need to prove that 	${}^{\overline{\underline{\mathbf{I}}}} \iota$ and ${}^{\overline{\underline{\mathbf{I}}}} \iota ^{-1}$ are bijective and their respective inverse function.
	
	${}^{\overline{\underline{\mathbf{I}}}} \iota$ and ${}^{\overline{\underline{\mathbf{I}}}} \iota ^{-1}$ are bijective and their respective inverse function if and only if
	\begin{gather}
		\label{eq:proof:translation_01}\forall j\in \left\{1,...,\prod _{\delta =1}^{d} I_{\delta }\right\} :{}^{\overline{\underline{\mathbf{I}}}} \iota ^{-1}\left( \ ^{\overline{\underline{\mathbf{I}}}} \iota (j)\right) =j\\
		\label{eq:proof:translation_02}\land \quad  \forall \underline{j} \in \mathbb{N}^{d} \cap \left[\overline{\underline{\mathbf{1}}} ,\overline{\underline{\mathbf{I}}}\right] :{}^{\overline{\underline{\mathbf{I}}}} \iota \left( \ ^{\overline{\underline{\mathbf{I}}}} \iota ^{-1} (\underline{j} )\right) =\underline{j}
	\end{gather}
	
	First, we proof equation \ref{eq:proof:translation_01}:
	\begin{align}
		^{\overline{\underline{\mathbf{I}}}} \iota ^{-1}\left( \ ^{\overline{\underline{\mathbf{I}}}} \iota (j)\ \right) & =1+\left(\left(( j-1) -\left\lfloor \frac{j-1}{I_{1}}\right\rfloor I_{1} +1\right) -1\right)\
		\ \ +\left(\left(\left\lfloor \frac{j-1}{I_{1}}\right\rfloor -\left\lfloor \frac{j-1}{I_{1} I_{2}}\right\rfloor I_{2} +1\right) -1\right) I_{1}\\
		& \ \ +\left(\left(\left\lfloor \frac{j-1}{I_{1} I_{2}}\right\rfloor -\left\lfloor \frac{j-1}{I_{1} I_{2} I_{3}}\right\rfloor I_{3} +1\right) -1\right) I_{1} I_{2}\\
		&\ \ +...+\left(\left(\left\lfloor \frac{j-1}{\prod _{t=1}^{l-1} I_{t}}\right\rfloor -\left\lfloor \frac{j-1}{\prod _{t=1}^{l} I_{t}}\right\rfloor I_{l} +1\right) -1\right)\prod _{t=1}^{l-1} I_{t}\\
		& \ \ +\left(\left(\left\lfloor \frac{j-1}{\prod _{t=1}^{l} I_{t}}\right\rfloor -\left\lfloor \frac{j-1}{\prod _{t=1}^{l+1} I_{t}}\right\rfloor I_{l+1} +1\right) -1\right)\prod _{t=1}^{l} I_{t}\\
		& \ \ +...+\left(\left(\left\lfloor \frac{j-1}{\prod _{t=1}^{d-2} I_{t}}\right\rfloor -\left\lfloor \frac{j-1}{\prod _{t=1}^{d-1} I_{t}}\right\rfloor I_{d-1} +1\right) -1\right)\prod _{t=1}^{d-2} I_{t}\\
		&\ \ +\left(\left(\left\lfloor \frac{j-1}{\prod _{t=1}^{d-1} I_{t}}\right\rfloor +1\right) -1\right)\prod _{t=0}^{d-1} I_{t}
	\end{align}
	\begin{align}
		& =1+( j-1) \ \ \underbrace{-\left\lfloor \frac{j-1}{I_{1}}\right\rfloor I_{1} +\left\lfloor \frac{j-1}{I_{1}}\right\rfloor I_{1}}_{=0}\
		\ \ \underbrace{-\left\lfloor \frac{j-1}{I_{1} I_{2}}\right\rfloor I_{2} I_{1} +\left\lfloor \frac{j-1}{I_{1} I_{2}}\right\rfloor I_{1} I_{2}}_{=0}\\
		& \ \ -\left\lfloor \frac{j-1}{I_{1} I_{2} I_{3}}\right\rfloor I_{3} I_{1} I_{2} +...+\left\lfloor \frac{j-1}{\prod _{t=1}^{l-1} I_{t}}\right\rfloor \prod _{t=1}^{l-1} I_{t}\
		\ \ \underbrace{-\left\lfloor \frac{j-1}{\prod _{t=1}^{l} I_{t}}\right\rfloor \prod _{t=1}^{l} I_{t} +\left\lfloor \frac{j-1}{\prod _{t=1}^{l} I_{t}}\right\rfloor \prod _{t=1}^{l} I_{t}}_{=0}\\
		& \ \ -\left\lfloor \frac{j-1}{\prod _{t=1}^{l+1} I_{t}}\right\rfloor \prod _{t=1}^{l+1} I_{t} +...+\left\lfloor \frac{j-1}{\prod _{t=1}^{d-2} I_{t}}\right\rfloor \prod _{t=1}^{d-2} I_{t}\
		\ \ \underbrace{-\left\lfloor \frac{j-1}{\prod _{t=1}^{d-1} I_{t}}\right\rfloor I_{d-1}\prod _{t=1}^{d-2} I_{t} +\left\lfloor \frac{j-1}{\prod _{t=1}^{d-1} I_{t}}\right\rfloor \prod _{t=1}^{d-1} I_{t}}_{=0}\\
		& \underline{=j}
	\end{align}
	
	Second. we proof equation \ref{eq:proof:translation_02} by subdividing the resulting vector in its entries. Starting with the first entry
	\begin{align}
		\left< ^{\overline{\underline{\mathbf{I}}}} \iota \left(^{\overline{\underline{\mathbf{I}}}} \iota ^{-1} (\ \underline{j} )\ \right)\right> _{1} = & \ \left(\left( 1+( j_{1} -1) +( j_{2} -1) I_{1} +...+( j_{d} -1)\prod _{t=1}^{d-1} I_{t}\right) -1\right)\\
		& -\left\lfloor \frac{\left( 1+( j_{1} -1) +...+( j_{d} -1)\prod _{t=1}^{d-1} I_{t}\right) -1}{I_{1}}\right\rfloor I_{1} +1\\
		( side\ calculations) & \left[\begin{array}{ r l }
			& \left\lfloor \frac{\left( 1+( j_{1} -1) +...+( j_{d} -1)\prod _{t=1}^{d-1} I_{t}\right) -1}{I_{1}}\right\rfloor I_{1}\\
			= & \left\lfloor \frac{j_{1} -1}{I_{1}} +\frac{( j_{2} -1) I_{1} +...+( j_{d} -1)\prod _{t=1}^{d-1} I_{t}}{I_{1}}\right\rfloor I_{1}\\
			= & \left\lfloor \underbrace{\frac{j_{1} -1}{I_{1}}}_{1\leq j_{1} \leq I_{1}} +\underbrace{( j_{2} -1) +...+( j_{d} -1)\prod _{t=2}^{d-1} I_{t}}_{\in \mathbb{N}_{0}}\right\rfloor I_{1}\\
			= & \left(( j_{2} -1) +...+( j_{d} -1)\prod _{t=2}^{d-1} I_{t}\right) I_{1}\\
			= & ( j_{2} -1) I_{1} +...+( j_{d} -1)\prod _{t=1}^{d-1} I_{t}
		\end{array}\right]\\
		= & ( j_{1} -1) +( j_{2} -1) I_{1} +...+( j_{d} -1)\prod _{t=1}^{d-1} I_{t}\ 
		-\left(( j_{2} -1) I_{1} +...+( j_{d} -1)\prod _{t=1}^{d-1} I_{t}\right) +1\\
		= & j_{1}.
	\end{align}
	We proceed with all entries except the first and last entry:

	\begin{align}
		\forall l\in \{2,...,d-1\}:\\
		\left< ^{\overline{\underline{\mathbf{I}}}} \iota \left(^{\overline{\underline{\mathbf{I}}}} \iota ^{-1} (\ \underline{j} )\ \right)\right> _{l}\
		&=  \ \left\lfloor \frac{\left( 1+( j_{1} -1) +...+( j_{d} -1)\prod _{t=1}^{d-1} I_{t}\right) -1}{\prod _{t=1}^{l-1} I_{t}}\right\rfloor \\
		& -\left\lfloor \frac{\left( 1+( j_{1} -1) +...+( j_{d} -1)\prod _{t=1}^{d-1} I_{t}\right) -1}{\prod _{t=1}^{l} I_{t}}\right\rfloor I_{l} +1\\
		& auxiliary\ calculations:\\
		& \left[\begin{array}{ r l }
			& \left\lfloor \frac{\left( 1+( j_{1} -1) +...+( j_{d} -1)\prod _{t=1}^{d-1} I_{t}\right) -1}{\prod _{t=1}^{l-1} I_{t}}\right\rfloor \\
			= & \left\lfloor \frac{( j_{1} -1) +...+( j_{l-1} -1)\prod _{t=1}^{l-2} I_{t}}{\prod _{t=1}^{l-1} I_{t}} +\frac{( j_{l} -1)\prod _{t=1}^{l-1} I_{t} +...+( j_{d} -1)\prod _{t=1}^{d-1} I_{t}}{\prod _{t=1}^{l-1} I_{t}}\right\rfloor \\
			& auxiliary\ calculations:\\
			& \left[\begin{array}{ r l }
				0\leq  & \underbrace{( j_{1} -1)}_{1\leq j_{1} \leq I_{1}} +...+\underbrace{( j_{l-1} -1)}_{1\leq j_{l-1} \leq I_{l-1}}\prod _{t=1}^{l-2} I_{t}\\
				\leq  & I_{1} -1+(I_{2} -1)I_{1} +...+(I_{l-1} -1)\prod _{t=1}^{l-2} I_{t}\\
				= & I_{1} -1+I_{2} I_{1} -I_{1} +...+\prod _{t=1}^{l-1} I_{t} -\prod _{t=1}^{l-2} I_{t}\\
				= & \prod _{t=0}^{l-1} I_{t} -1\\
				\prod _{t=0}^{l-1} I_{t} -1<  & \prod _{t=0}^{l-1} I_{t}
			\end{array}\right]\\
			= & \left\lfloor \underbrace{\frac{( j_{1} -1) +...+( j_{l-1} -1)\prod _{t=1}^{l-2} I_{t}}{\prod _{t=1}^{l-1} I_{t}}}_{\geq 0,\ < 1} +\underbrace{( j_{l} -1) +...+( j_{d} -1)\prod _{t=l}^{d-1} I_{t}}_{\in \mathbb{N}_{0}}\right\rfloor \\
			= & ( j_{l} -1) +...+( j_{d} -1)\prod _{t=l}^{d-1} I_{t}
		\end{array}\right]\\
		= & \left(( j_{l} -1) +...+( j_{d} -1)\prod _{t=l}^{d-1} I_{t}\right) -\left(( j_{l+1} -1) +...+( j_{d} -1)\prod _{t=l+1}^{d-1} I_{t}\right) I_{l} +1\\
		= & ( j_{l} -1) +( j_{l+1} -1) I_{l} +...+( j_{d} -1)\prod _{t=l}^{d-1} I_{t} -( j_{l+1} -1) I_{l} -...-( j_{d} -1)\prod _{t=l}^{d-1} I_{t} +1\\
		= & j_{l}
	\end{align}
	
	We finish with the last entry:
	\begin{align}
		\left< \ ^{\overline{\underline{\mathbf{I}}}} \iota \left( \ ^{\overline{\underline{\mathbf{I}}}} \iota ^{-1} (\ \underline{j} )\ \right)\right> _{d} = & \ \left\lfloor \frac{\left( 1+( j_{1} -1) +...+( j_{d} -1)\prod _{t=1}^{d-1} I_{t}\right) -1}{\prod _{t=1}^{d-1} I_{t}}\right\rfloor +1\\
		= & \left\lfloor \underbrace{\frac{1+( j_{1} -1) +...+( j_{d-1} -1)\prod _{t=1}^{d-2} I_{t}}{\prod _{t=1}^{d-1} I_{t}}}_{< 1} +\underbrace{( j_{d} -1)}_{\in \mathbb{N}_{0}}\right\rfloor +1\\
		= & j_{d}.
	\end{align}

	Taking these three results together, we can conclude
	\begin{align}
		^{\overline{\underline{\mathbf{I}}}} \iota \left(^{\overline{\underline{\mathbf{I}}}} \iota ^{-1}( \ \underline{j}) \ \right) & =\left[\begin{array}{ c }
			j_{1}\\
			j_{2}\\
			\vdots \\
			j_{d}
		\end{array}\right]\\
		& \underline{=\underline{j} \ }.
	\end{align}
	Hence, $^{\overline{\underline{\mathbf{I}}}} \iota$ and $^{\overline{\underline{\mathbf{I}}}} \iota ^{-1}$ are bijective and their respective inverse.
\end{proof}

\begin{proof} \textbf{of lemma \ref{lemma:OrderIdependenceOfInteractSeries}}\\
	We need to prove that the order independence of the interact function follows the order independence of a series of interactions.
	\begin{align}
		& _{1} i_{g}(_{1} i_{g} (p_{j} ,p_{k} ),p_{k'}) ={}_{1} i_{g} (_{1} i_{g} (p_{j} ,p_{k'} ),p_{k''} )\\
		\rightarrow  & \tilde{p} \ *_{_{1} i_{g}} \ \sigma ( p_{1} ,...,p_{n}) =\tilde{p} \ *_{_{1} i_{g}} \ ( p_{1} ,...,p_{n})
	\end{align}
	
	The proof idea is to use the bubble sort strategy to go from an arbitrary permutation $\sigma$ of the interacting particles to the original order. Therefore, we prove that the necessary swap of two consecutive particles does not change the result.
	\begin{align}
		_{1} i_{g} (_{1} i_{g} (p_{j} ,p_{k} ),p_{k'} )&={}_{1} i_{g} (_{1} i_{g} (p_{j} ,p_{k'} ),p_{k''} )\\
		\rightarrow   \tilde{p} \ *_{{}_{1} i_{g}} \ \sigma (p_{1} ,...,p_{n} )\
		&=\underbrace{\tilde{p} \ *_{{}_{1} i_{g}} \ (\sigma (p_{1} ),...,\sigma (p_{j-1} )}_{=:p'} ,\sigma (p_{j} ),\sigma (p_{j+1} ),...,\sigma (p_{n} ))\\
		& =p'\ *_{_{1} i_{g}} \ ( \sigma (p_{j} ),\sigma (p_{j+1} ),...,\sigma (p_{n} ))\\
		& ={}_{1} i_{g}(_{1} i_{g} (p',\sigma (p_{j} )),\sigma (p_{j+1} )) \ *_{_{1} i_{g}} \ ( \sigma (p_{j+2} ),...,\sigma (p_{n} ))\\
		& ={}_{1} i_{g}(_{1} i_{g} (p',\sigma (p_{j+1} )),\sigma (p_{j} )) \ *_{_{1} i_{g}} \ ( \sigma (p_{j+2} ),...,\sigma (p_{n} ))\\
		& =\tilde{p} \ *_{_{1} i_{g}} \ ( \sigma (p_{1} ),...,\sigma (p_{j-1} ),\sigma (p_{j+1} ),\sigma (p_{j} ),\sigma (p_{j+2} ),...,\sigma (p_{n} ))\\
		& \text{using bubble sort }\\
		& =\tilde{p} \ *_{_{1} i_{g}} \ (p_{1} ,...,p_{n} )
	\end{align}
\end{proof}
\newpage

\begin{proof} \textbf{of lemma \ref{lemma:copyAll:NoReadConflicts}}\\
	We need to poof that the function $copy_{\tilde{g}}^{ALL}$ does not induce overlapping communications. \\
	\\
	Overlapping lapping means that two processes communicate either to the same other third process or to each other.
	The potential overlapping communications are avoided by letting only some processes communicate simultaneously. The communicating processes are distributed in a checkerboard-like structure. Between two communicating processes are always two passive processes. Since each process communicates only with its direct neighbor processes, there can not be an overlapping communication. To prove that, we need to prove
	\begin{equation}
		\begin{array}{l}
			\forall k,l',l''\in \left\{1,...,3^{d}\right\} \ \forall j',j''\in \left\{1,..., {}^{k}N_{cell}^{*}  \right\} :\ \\
			j'\neq j''\rightarrow \ \beta ( \gamma ( k,j') ,l') \neq \beta ( \gamma ( k,j'') ,l'') \ \lor \  \beta ( \gamma ( k,j') ,l')=undef.\ ,
		\end{array}
	\end{equation}
	where $k$ is the number of the checkerboard-like pattern, $\gamma ( k,j)$ (eq. \ref{eq:definition:gamma}) is an index of a reading process and $\beta ( \gamma ( k,j) ,l)$ (eq. \ref{eq:definition:beta}) its $l$-th neighbor with it communicates. Hence, two communicating processes do not have a common process with which they communicate. If $\beta$ is undefined, there is no process.
	
	Inserting the definition of $\gamma$  into $\beta$  results in
	\begin{align}\label{eq:distributedpull:betagamma}
		\beta ( \gamma ( k,j) ,l) & ={}^{\ovun I} \iota ^{-1}\left(  {}^{\ovun{I}} \iota \left( {}^{\ovun{I}} \iota ^{-1}\left( \ {}^{{}^k\ovun{I}^{*}}\!\! \iota (j)\cdot 3+{}^{\ovun{3}} \iota (k)-\ovun{3}\right)\right)+{}^{\ovun{3}} \iota (l)-\ovun{2}\right)
		\\
		& ={}^{\ovun I} \iota ^{-1}\left( \ \ {}^{{}^{k}\ovun I^{*}} \iota (j)\cdot 3+{}^{\ovun 3} \iota (k)\ +{}^{\ovun 3} \iota (l)- \ovun 5 \right)
	\end{align}

	From here on, we do a proof by contradiction. Therefore, we negate the statement we want to prove and derive absurdity.
	\\
	
	Assuming 
	\begin{equation}
		j' \neq j''\ \land \ \beta ( \gamma ( k,j') ,l') = \beta ( \gamma ( k,j'') ,l'') \ \land \  \beta ( \gamma ( k,j') ,l')\neq undef.
	\end{equation}

	$\xleftrightarrow{ (\ref{eq:distributedpull:betagamma})}$
	\begin{equation}
		\begin{array}{ c }
			j' \neq j''
			\\ \land \	
			{}^{\ovun I} \iota ^{-1}\left(  {}^{{}^{k}\ovun I^{*}}\! \iota (j' )\cdot 3+{}^{\ovun 3} \iota (k) +{}^{\ovun 3} \iota (l')- \ovun 5 \right)
			={}^{\ovun I} \iota ^{-1}\left( {}^{{}^{k}\ovun I^{*}}\! \iota (j'')\cdot 3+{}^{\ovun 3} \iota (k) +{}^{\ovun 3} \iota (l'')- \ovun 5 \right)
			\\ \land \
			{}^{\ovun I} \iota ^{-1}\left(  {}^{{}^{k}\ovun I^{*}}\! \iota (j')\cdot 3+{}^{\ovun 3} \iota (k) +{}^{\ovun 3} \iota (l')- \ovun 5 \right)\neq undef.
		\end{array}
	\end{equation}

	$\xleftrightarrow{lemma\  \ref{lemma:indexTransformIsBijectiv}}$
	\begin{equation}
		\begin{array}{ c }
			j' \neq j''
			\\ \land \	\Bigg(
			{}^{{}^{k}\ovun I^{*}}\! \iota (j' )\cdot 3+{}^{\ovun 3} \iota (k) +{}^{\ovun 3} \iota (l')- \ovun 5 
			= {}^{{}^{k}\ovun I^{*}}\! \iota (j'')\cdot 3+{}^{\ovun 3} \iota (k) +{}^{\ovun 3} \iota (l'')- \ovun 5 
			\\ \lor \ 
			{}^{\ovun I} \iota ^{-1}\left(  {}^{{}^{k}\ovun I^{*}}\! \iota (j')\cdot 3+{}^{\ovun 3} \iota (k) +{}^{\ovun 3} \iota (l')- \ovun 5 \right)= undef.
			\Bigg) \\ \land \
			{}^{\ovun I} \iota ^{-1}\left(  {}^{{}^{k}\ovun I^{*}}\! \iota (j')\cdot 3+{}^{\ovun 3} \iota (k) +{}^{\ovun 3} \iota (l')- \ovun 5 \right)\neq undef.
		\end{array}
	\end{equation}
	
	$\longleftrightarrow$
	\begin{equation}
		\begin{array}{ c }
			j' \neq j''
			\\ \land \
			{}^{{}^{k}\ovun I^{*}}\! \iota (j' )\cdot 3+{}^{\ovun 3} \iota (k) +{}^{\ovun 3} \iota (l')- \ovun 5 
			= {}^{{}^{k}\ovun I^{*}}\! \iota (j'')\cdot 3+{}^{\ovun 3} \iota (k) +{}^{\ovun 3} \iota (l'')- \ovun 5 
			\\ \land \
			{}^{\ovun I} \iota ^{-1}\left(  {}^{{}^{k}\ovun I^{*}}\! \iota (j')\cdot 3+{}^{\ovun 3} \iota (k) +{}^{\ovun 3} \iota (l')- \ovun 5 \right)\neq undef.
		\end{array}
	\end{equation}
	
	$\xleftrightarrow{def.\ \ref{def:indexTransform}}$
	\begin{equation}
			j' \neq j''
			\ \land \ \
			\underbrace{{}^{{}^{k}\ovun I^{*}}\! \iota (j' )}_{\in \mathbb{N}_{1}^{d}}
			\cdot 3 +
			\underbrace{{}^{\ovun 3} \iota (l')}_{\in\left[\ovun{1},\ovun{3}\right]} 
			= \underbrace{{}^{{}^{k}\ovun I^{*}}\! \iota (j'')}_{\in \mathbb{N}_{1}^{d}}
			\cdot 3 +
			\underbrace{{}^{\ovun 3} \iota (l'')}_{\in\left[\ovun{1},\ovun{3}\right]}   
			\ \land^{\myhh}  \ \
			{}^{{}^{k}\ovun I^{*}}\! \iota (j')\cdot 3+{}^{\ovun 3} \iota (k) +{}^{\ovun 3} \iota (l')- \ovun 5 \in \mathbb{N}^d_1\cap\left[\ovun{1},\ovun{I}\right]
	\end{equation}

	$\longleftrightarrow$
	\begin{equation}
			j' \neq j''
			\ \ \land \ \
			\underbrace{{}^{{}^{k}\ovun I^{*}}\! \iota(j')
				-{}^{{}^{k}\ovun I^{*}}\! \iota(j'')}_{\in \mathbb{Z}^{d}}
			\cdot 3 
			=\underbrace{{}^{\ovun 3} \iota(l'')-{}^{\ovun 3} \iota (l')}_{\in\left[-\ovun{2},\ovun{2}\right]}   
			\ \ \land^{\myhh} \ \ 
			{}^{{}^{k}\ovun I^{*}}\! \iota (j')\cdot 3+{}^{\ovun 3} \iota (k) +{}^{\ovun 3} \iota (l')- \ovun 5 \in \mathbb{N}^d_1\cap\left[\ovun{1},\ovun{I}\right]
	\end{equation}
	
	$\xleftrightarrow{def.\ \ref{def:indexTransform}}$
	\begin{equation}
			j' \neq j''
			\ \ \land \ \
			\underbrace{{}^{{}^{k}\ovun I^{*}}\! \iota(j')
				-{}^{{}^{k}\ovun I^{*}}\! \iota(j'')}_{\in \mathbb{Z}^{d}}
			\cdot 3 
			=\underbrace{{}^{\ovun 3} \iota(l'')-{}^{\ovun 3} \iota (l')}_{\in\left[-\ovun{2},\ovun{2}\right]} 
			= \ovun 0  
			\ \ \land^{\myhh} \ \
			{}^{{}^{k}\ovun I^{*}}\! \iota (j')\cdot 3+{}^{\ovun 3} \iota (k) +{}^{\ovun 3} \iota (l')- \ovun 5 \in \mathbb{N}^d_1\cap\left[\ovun{1},\ovun{I}\right]
	\end{equation}
	
	$\xleftrightarrow{lemma\ \ref{lemma:indexTransformIsBijectiv}}$
	\begin{equation}
			j' \neq j''
			\ \ \land \ \
			j'=j'' \ \land \ l'=l''
			\ \ \land^{\myhh} \ \
			{}^{{}^{k}\ovun I^{*}}\! \iota (j')\cdot 3+{}^{\ovun 3} \iota (k) +{}^{\ovun 3} \iota (l')- \ovun 5 \in \mathbb{N}^d_1\cap\left[\ovun{1},\ovun{I}\right]
	\end{equation}
	\begin{equation}
		\lightning (\text{contradiction})
	\end{equation}
	
	$\longrightarrow$
	\begin{equation}
		\begin{array}{l}
			\forall k,l',l''\in \left\{1,...,3^{d}\right\} \ \forall j',j''\in \left\{1,..., {}^{k}N_{cell}^{*}  \right\} :\ \\
			j'\neq j''\rightarrow \ \beta ( \gamma ( k,j') ,l') \neq \beta ( \gamma ( k,j'') ,l'') \ \lor \  \beta ( \gamma ( k,j') ,l')=undef.\ ,
		\end{array}
	\end{equation}
\end{proof}

\begin{proof}\textbf{of lemma \ref{lemma:copyAll:AllNeighborsAreThere}}\\
	We need to prove, after the $copy^{ALL}_{\tilde{g}}$ function, the storages of all processes contain the particles of the corresponding cells and all their neighbor particles.
	Hence, first, we need to prove that each process executes the $copy^{ALL}_{\tilde{g}}$ function and second, that on a process after the $copy^{ALL}_{\tilde{g}}$ function all neighbor particles of all particles of the center storage compartment are in the storage.
	
	To first, the idea is to prove that $\gamma(k,j)$ is bijective and the codomain is the set of all indices of the processes $\{1,...,N_{cell}\}$,
	
	\begin{equation}
		\gamma :\left\{1,...,3^{d}\right\} \times \left\{1,...,{}^{k}N_{cell}^{*}  \right\}\rightarrow \left\{1,...,N_{cell}\right\} \ \text{is bijective}.
	\end{equation}
	We define a helper function $\underline \gamma^*$, prove it is bijective and transform it into $\gamma$.\\
	We declare
	\begin{equation}
		\underline \gamma ^{*} :\{1,...,3\}^{d} \times \mathbb{N}^{d} \cap \left[\ovun 1 ,\ ^{k}\ovun I^{*}\right]\rightarrow \mathbb{N}^{d} \cap \left[\ovun 1 ,\ovun I \right] 
	\end{equation}
	and define
	\begin{equation}
		\underline \gamma ^{*}(\underline{k} ,\underline{j}):=\underline{k} + (\underline{j}-\ovun 1) \cdot 3.
	\end{equation}
	One element of $\underline \gamma ^{*}$ is $\left< \underline \gamma ^{*}(\underline{k} ,\underline{j})\right> _{w}$ and we can rewrite it by 
	\begin{equation}
		\left<\underline \gamma ^{*}(\underline{k} ,\underline{j})\right> _{w} =1+(k_{w}-1) +(j_{w}-1) \cdot 3=\underbrace{^{^{k}\underline{\overline{\mathbf{J}}}_{w}} \iota ^{-1}\left(\left(\begin{array}{ c }
				k_{w}\\
				j_{w}
			\end{array}\right)\right)}_{\text{is bijective} \ \left( \iota \ \text{is bijective}\right)} 
	\end{equation}
	where ${}^{k}\ovun J_{w} :=\left(\begin{array}{ c }
		3\\
		^{k} I_{w}^{*}
	\end{array}\right)$.
	The domain of the arguments $j$ and $k$ depend on each other. 
	Hence, it is not obvious that $\underline\gamma^*$ is not leaving the codomain $\mathbb{N}^{d} \cap \left[\ovun 1 ,\ovun I \right]$. $\underline\gamma^*$ is monotone. Therefore, it is sufficient to prove that $\underline\gamma^*$ is in the codomain for the smallest and largest arguments. 
	The minimal value of $\underline\gamma^*$ is
	\begin{equation}
		min\left(\left<\underline \gamma ^{*}(\underline{k} ,\underline{j})\right> _{w}\right) =1+(1-1)\cdot 3=\underline{1}.
	\end{equation}
	This is in the codomain. 
	The maximal value of $\underline\gamma^*$ is
	\begin{equation}
		max\left(\left< \underline\gamma ^{*}(\underline{k} ,\underline{j})\right> _{w}\right) =k_{w}^{max} +\ \left( \ ^{k^{max}} I_{w}^{*} -1\right) \cdot 3.
	\end{equation}
	Using the definition of ${}^{k_{max}}\ovun I^*$ (eq. \ref{eq:definition:kIstar}) leads to
	\begin{equation}
		max\left(\left<\underline \gamma ^{*}(\underline{k} ,\underline{j})\right> _{w}\right)
		=k_{w}^{max} +\ \left( \ \left\lfloor \tfrac{1}{3}\left( I_{w} -k_{w}^{max} +3\right)\right\rfloor -1\right) \cdot 3
	\end{equation}
	We substituted 
	\begin{equation}
		I_{w} -k_{w}^{max} +3=:\underbrace{T'}_{\in \mathbb N}\cdot 3+\underbrace{T''}_{\in \{0,1,2\}}
	\end{equation}
	and get
	\begin{align}
		max\left(\left<\underline \gamma ^{*}(\underline{k} ,\underline{j})\right> _{w}\right)
		&=k_{w}^{max} +\left(\left\lfloor \tfrac{3T'}{3} +\tfrac{T''}{3}\right\rfloor -1\right) \cdot 3
		\\
		&=k_{w}^{max} +( T'-1) \cdot 3.
	\end{align}
	We rearrange the substitution to
	\begin{equation}
		3T'=I_{w} -k_{w}^{max} +3-T''
	\end{equation}
	and plug it in
	\begin{align}
		max\left(\left<\underline \gamma ^{*}(\underline{k} ,\underline{j})\right> _{w}\right)
		&=k_{w}^{max} +I_{w} -k_{w}^{max} +3-T''-3
		\\
		&=I_{w}-\underbrace{T''}_{\in \{0,1,2\}}.
	\end{align}
	This means
	\begin{equation}
		max\left(\left<\underline \gamma ^{*}(\underline{k} ,\underline{j})\right> _{w}\right)
		\leq I_w.
	\end{equation}
	It remains to be shown that $I_w$ can be reached. This is only possible if $T''=0$. Hence, we need to prove
	\begin{equation}
		\exists k_{w} \in \{1,2,3\} :T''=0.
	\end{equation}
	From $T''=0$ follows that
	\begin{equation}
		\frac{3T'+T''}{3} =\frac{I_{w} +3-k_{w}^{max}}{3}=n \in \mathbb{N}.
	\end{equation}
	
	$\longrightarrow$
	\begin{equation}\label{eq:gammabijective:kMax01}
		n+\frac{k_{w}^{max}-1}{3}=\frac{I_{w} +2}{3}
	\end{equation}
	Taking the $floor$ on both sides lead to
	\begin{equation}
		\left\lfloor n+\underbrace{\frac{k_{w}^{max}-1}{3}}_{\in \{0,1,2\}}\right\rfloor=\left\lfloor\frac{I_{w} +2}{3}\right\rfloor
	\end{equation}
	
	$\longrightarrow$
	\begin{equation}
		n=\left\lfloor\frac{I_{w} +2}{3}\right\rfloor.
	\end{equation}
	Inserting this into equation \ref{eq:gammabijective:kMax01} lead to
	\begin{equation}
		k_{w} =( (I_{w} +3) -1 ) -\left\lfloor \tfrac{(I_{w} +3)-1}{3}\right\rfloor \cdot 3 +1 = \left< {}^{\ovun J} \iota ( I_{w} +3)\right> _{1} \in \{1,2,3\}
	\end{equation}
	where $\ovun{J} = (3,...)^{\mathbf{T}}$. We can follow,
	\begin{equation}
		\forall w\in \{1,...,d\} :\left< \underline\gamma ^{*}\right> _{w} \ \text{is bijective}
	\end{equation}
	
	$\longrightarrow$
	\begin{equation}
		\underline\gamma ^{*} \ \text{is bijective}
	\end{equation}
	Now we need to transform $\underline\gamma^*$ to $\gamma$.
	We substitute  $\underline k$ and $\underline j$ by
	\begin{equation}
		\underline k = {}^{\ovun 3}\iota(k), \qquad \underline j= {}^{{}^{k}\ovun I^*}\iota (j).
	\end{equation}
	We know that $\iota$ is bijective (lemma \ref{lemma:indexTransformIsBijectiv}, def. \ref{def:indexTransform}) and therefore  
	\begin{equation}
		k\in \{1,...,3^d\}, \qquad j\in\mathbb~N^d\cap\left[\ovun{1},{}^{k}\ovun{I}^*\right].
	\end{equation} 
	We also know $\iota^{-1}$ is bijective. Hence, 
	\begin{align}
		{}^{\ovun I}\iota^{-1}\left(\underline \gamma^*\left({}^{\ovun 3}\iota(k), {}^{{}^{k}\ovun I^*}\iota (j)\right)\right)
		&={}^{\ovun I}\iota^{-1}\left({}^{\ovun 3}\iota(k)+ \left({}^{{}^{k}\ovun I^*}\iota (j)-\ovun 1\right)\cdot 3\right)
		\\
		&=\gamma(k,j)
	\end{align}
	This means $\gamma$ is bijective, and the $copy^{ALL}_{\tilde{g}}$ function is executed for all processes.
	\\[12pt]
	To second,
	it remains to be proven that the neighbor particles of all particles of the center storage compartment are in the storage.
	The function $copy_{\left( \tilde{g} ,\mathcal P ,w\right)}(\underline{\mathbf{p}} ,l)$ (eq. \ref{eq:copy:fromOneCPU}) copies the central $\left(\tfrac{3^d+1}{2}\right)$ storage compartment form the $\beta(w,l)$-th process to the $l$-th storage compartment on the $w$-th process.
	We start from the vectorial index view to find the corresponding cell/process where the particles are for the $l$-th storage compartment on the $w$-th process. 
	The vectorial index of the $w$-th process is ${}^{\ovun I}\iota(w)$ and the vectorial index of $l$-th storage compartment is ${}^{\ovun 3}\iota(l)\in \{1,2,3\}$. The storage center compartment $l=\left(\tfrac{3^d+1}{2}\right)$ corresponds to the cell belonging to the process. The rest of the storage compartments should represent the surrounding cells.
	Therefore, we need to shift the storage compartment index by $-2$ in all dimensions to account for this. Hence, the vectorial index of the corresponding process for the $l$-th storage compartment of the $w$-th process is
	\begin{equation}
		{}^{\ovun I}\iota(w)+{}^{\ovun 3}\iota(l)-\ovun 2
	\end{equation}
	Transforming the vectorial index to a scalar index results in
	\begin{equation}
		{}^{\ovun I}\iota^{-1}\left({}^{\ovun I}\iota(w)+{}^{\ovun 3}\iota(l)-\ovun 2\right)=\beta(w,l).
	\end{equation}
	Hence, the $\beta(w,l)$-th process has the corresponding particles in its center storage compartment. The storage gets the particles from the surrounding cells by the $copy$ function from the other processes. With this, we can derive the domain area that the storage covers.
	The vectorial indices of the cells that belong to the domain are
	\begin{equation}\label{eq:copy:allStorageIndeces}
		\left\{{}^{\ovun I}\iota(\beta(w,l)): l\in\{1,...,3^d\}\right\} = \left[{}^{\ovun I}\iota(w)-\ovun{1},\ {}^{\ovun I}\iota(w)+\ovun{1}\right]\cap \left[\ovun 1, \ovun{I}\right]\cap \mathbb{N}^d.
	\end{equation}
	Per definition (eq. \ref{eq:initialCell:ParticleDistribution}) the particle that belong to the $w$-th cell  are
	\begin{equation}
		p_j=	(...,x_j,...) \in \mathbf{p}^1_{w}:\  {}^{\ovun{I}}\iota(w)= \left \lfloor \frac{1}{r_c} ( \underline{x}_j-\underline{D}_{min})\right \rfloor + \ovun{1}.
	\end{equation}
	
	$\longrightarrow$
	\begin{equation}
		p_j \in \mathbf{p}^1_{w}:\  {}^{\ovun{I}}\iota(w)-\ovun{1}= \left \lfloor \frac{1}{r_c} ( \underline{x}_j-\underline{D}_{min})\right \rfloor .
	\end{equation}
	
	$\longrightarrow$
	\begin{equation}
		p_j \in \mathbf{p}^1_{w}:\  \frac{1}{r_c} ( \underline{x}_j-\underline{D}_{min}) \in \left[{}^{\ovun{I}}\iota(w)-\ovun{1},\ {}^{\ovun{I}}\iota(w)-\ovun{1} +\ovun{1}\right) .
	\end{equation}
	
	$\longrightarrow$
	\begin{equation}\label{eq:copyproof:CellSpace}
		p_j \in \mathbf{p}^1_{w}:\  
		\underline{x}_j \in \left[
		\left({}^{\ovun{I}}\iota(w)-\ovun{1}\right)\cdot r_c+\underline{D}_{min},\ 
		{}^{\ovun{I}}\iota(w)\cdot r_c+\underline{D}_{min} \right) .
	\end{equation}
	Taking all cells/compartments (eq. \ref{eq:copy:allStorageIndeces}) of the storage of the $w$-process leads to
	\begin{equation}
		p_j \in \underset{l=1}{\overset{3^{d}}{\fullmoon }} \langle {}^{[\text{PROC}w]}\underline{\mathbf{p} }^1 \rangle _{l}:\ 
		\underline{x}_j \in \left[
		\left(\left({}^{\ovun{I}}\iota(w)-\ovun{1}\right)-\ovun{1}\right)\cdot r_c+\underline{D}_{min},\ 
		\left({}^{\ovun{I}}\iota(w)+\ovun{1}\right)\cdot r_c+\underline{D}_{min} \right)\
		\cap \left[\underline{D}_{min}, \underline{D}_{max}\right),
	\end{equation}
	where $\left[\ovun 1, \ovun{I}\right]$ translates to $\left[\underline{D}_{min}, \underline{D}_{max}\right)$ (eq. \ref{eq:definitionOfI}).
	
	$\longrightarrow$
	\begin{equation}\label{eq:copyproof:finelsizeStorage}
		p_j \in \underset{l=1}{\overset{3^{d}}{\fullmoon }} \langle {}^{[\text{PROC}w]}\underline{\mathbf{p}}^1 \rangle _{l}:\  
		\underline{x}_j \in \left[
		\left({}^{\ovun{I}}\iota(w)-\ovun{2}\right)\cdot r_c+\underline{D}_{min},\ 
		\left({}^{\ovun{I}}\iota(w)+\ovun{1}\right)\cdot r_c+\underline{D}_{min} \right)
		\cap \left[\underline{D}_{min}, \underline{D}_{max}\right).
	\end{equation}
	It remains to be proven that the neighbor particles of all particles of the central storage compartment of the $w$-th process are in the storage. Hence, the domain area covered by the storage includes the area covered by the neighborhood function $u$.
	Be $p_k\in \mathbf{p}^1$ and $p_j \in \left<\mathbf{p}^1\right>_{u(g,\mathbf{p}^1,k)}$ (eq. \ref{eq:constraint:neighborhood}) then
	\begin{equation}
		\vert\underline{x}_k-\underline{x}_j\vert\leq r_c
	\end{equation}
	
	$\longrightarrow$
	\begin{equation}
		\underline{x}_j \in  \left[\underline{x}_k- \ovun 1 \cdot r_c,\ \underline{x}_k+ \ovun 1 \cdot r_c\right]
	\end{equation}
	Be now $p_k\in \mathbf{p}^1_w$ then we know (eq. \ref{eq:copyproof:CellSpace})
	\begin{equation}
		\underline{x}_k \in \left[
		\left({}^{\ovun{I}}\iota(w)-\ovun{1}\right)\cdot r_c+\underline{D}_{min},\ 
		{}^{\ovun{I}}\iota(w)\cdot r_c+\underline{D}_{min} \right) .
	\end{equation}
	
	$\longrightarrow$
	\begin{equation}
		\underline{x}_j \in \left[
		\left({}^{\ovun{I}}\iota(w)-\ovun{1}\right)\cdot r_c+\underline{D}_{min}-\ovun{1}\cdot r_c,\ 
		{}^{\ovun{I}}\iota(w)\cdot r_c+\underline{D}_{min}+\ovun{1}\cdot r_c \right)\
		\cap \left[\underline{D}_{min}, \underline{D}_{max}\right).
	\end{equation}
	
	$\longrightarrow$
	\begin{equation}
		\underline{x}_j \in \left[
		\left({}^{\ovun{I}}\iota(w)-\ovun{2}\right)\cdot r_c+\underline{D}_{min},\ 
		\left({}^{\ovun{I}}\iota(w)+\ovun{1}\right)\cdot r_c+\underline{D}_{min} \right)
		\cap \left[\underline{D}_{min}, \underline{D}_{max}\right).
	\end{equation}
	This is the same as equation \ref{eq:copyproof:finelsizeStorage}, which means that the neighbor particles of all particles of the center storage compartment of the $w$-th process are all in the storage of the $w$-th process, and this is true for all processes.
\end{proof}

\begin{proof} \textbf{of lemma \ref{lemma:distPlacesParticlesOnlyInsideStorages}}\\
	We need to prove that 
	the functions $dist_{\left(\tilde{g},\mathcal{G}^1\right)}^{ALL}$ places for each process the particles from the center storage compartments only to the other storage compartments of the same processes. Hence, it does not try to place them in a not existing storage compartment. Hence,
	\begin{multline}
		\forall w\in \{1,...,N_{cell}\}\forall p^1_j \in \left<\left<\mathcal{P}^1\right>_w\right>_{\tfrac{3^d+1}{2}}: 
		p^2_j:=  \left<\left<\left< 
		step_{\mathcal{G}^1}^{ALL}\left(copy_{\tilde{g}}^{ALL}(\mathcal{P}^1)\right)
		\right>_w\right>_{\tfrac{3^d+1}{2}}\right>_j\\
		\longrightarrow\ 
		\alpha =  {}^{\ovun 3} \iota^{-1}\left(\left\lfloor \tfrac{1}{r_{c}}(\underline{x}_j^2 -\underline{D}_{min})\right\rfloor - {}^{\ovun I} \iota ( w) +\ovun 3 \right)\in \{1,...,3^d\}.
	\end{multline}

	All particles have a position $\underline x \in \mathbb R^d$, hence, $p^1_j=(...,\underline x^1_j,...)$ and   $p^2_j=(...,\underline x^2_j,...)$.
	The condition (eq. \ref{eq:condition:MovmentIsSmallerThenRc}) restricts the movement of the particles to be smaller than $r_c$. Meaning for $\underline{\Delta x} := \underline x^2_j- \underline x^1_j$
	\begin{equation}
		\left|\underline{\Delta x} \right|\leq r_c.
	\end{equation}
	
	$\xrightarrow{def.\ \ref{def:vectorlength}}$
	\begin{equation}
		\left|\underline{\Delta x} \right| 
		= \left|\left(\begin{array}{ c }
			\Delta x _{1}\\
			\vdots \\
			\Delta x _{d}
		\end{array}\right)\right|
		=\sqrt{\Delta x^2 _{1}+...+\Delta x^2 _{d}}
		\leq r_c
	\end{equation}
	
	$\longrightarrow$
	\begin{equation}
		\forall k \in \{1,...,d\}: r_c \leq \Delta x_{k}
		\leq r_c
	\end{equation}
	
	$\longrightarrow$
	\begin{equation}
		\forall k \in \{1,...,d\}:\Delta x_{k}\in [ -r_c,r_c ]
	\end{equation}
	
	Using this, we can rewrite $\alpha$ from
	\begin{equation}
		\alpha =  {}^{\ovun 3} \iota^{-1}\left(\left\lfloor \tfrac{1}{r_{c}}(\underline{x}_j^2 -\underline{D}_{min})\right\rfloor - {}^{\ovun I} \iota ( w) +\ovun 3 \right)
	\end{equation}
	
	to
	\begin{equation}
		\alpha =  {}^{\ovun 3} \iota^{-1}\left(\left\lfloor \tfrac{1}{r_{c}}(\underline{x}_j^1+ \underline{\Delta x} -\underline{D}_{min})\right\rfloor - {}^{\ovun I} \iota ( w) +\ovun 3 \right).
	\end{equation}
	
	We put the index for the dimension as pre-sub-script.
	\begin{equation}
		\alpha =  {}^{\ovun 3} \iota^{-1}\left(
		\left(\begin{array}{ c }
			\left\lfloor \tfrac{1}{r_{c}}( {}_{1} x_j^1+ {}_{1}\Delta x - {}_{1}D_{min})\right\rfloor - \left<{}^{\ovun I} \iota ( w)\right>_1 +3\\
			\vdots \\
			\left\lfloor \tfrac{1}{r_{c}}( {}_{d} x_j^1+ {}_{d}\Delta x - {}_{d}D_{min})\right\rfloor - \left<{}^{\ovun I} \iota ( w)\right>_d +3
		\end{array}\right)
		\right)
	\end{equation}
	
	We know from the condition eq. \ref{eq:initialCell:ParticleDistribution} that
	\begin{equation}
		\forall p^1_j\in \mathbf p^1: p^1_j \in \mathbf{p}^1_{w}\ \text{with }\quad w= {}^{\ovun{I}}\iota^{-1}\left(\left \lfloor \frac{1}{r_c} ( \underline{x}^1_j-\underline{D}_{min})\right \rfloor + \ovun{1} \right).
	\end{equation}
	
	Taking this into $\alpha$ we get
	\begin{equation}
		\alpha =  {}^{\ovun 3} \iota^{-1}\left(
		\underbrace{\left(\begin{array}{ c }
				\underbrace{\Big\lfloor \tfrac{{}_{1} x_j^1 - {}_{1}D_{min}}{r_{c}} + \underbrace{\tfrac{{}_{1}\Delta x}{r_{c}}}_{\in [-1,1]} \Big\rfloor}_{\in \left\{\left<{}^{\ovun I} \iota ( w)\right>_1-2,\ \left<{}^{\ovun I} \iota ( w)\right>_1-1,\ \left<{}^{\ovun I} \iota ( w)\right>_1\right\}}
				- \left<{}^{\ovun I} \iota ( w)\right>_1 +3\\
				\vdots \\
				\underbrace{\Big\lfloor \tfrac{{}_{d} x_j^1 - {}_{d}D_{min}}{r_{c}} + \underbrace{\tfrac{{}_{d}\Delta x}{r_{c}}}_{\in [-1,1]} \Big\rfloor}_{\in \left\{\left<{}^{\ovun I} \iota ( w)\right>_d-2,\ \left<{}^{\ovun I} \iota ( w)\right>_d-1,\ \left<{}^{\ovun I} \iota ( w)\right>_d\right\}} 
				- \left<{}^{\ovun I} \iota ( w)\right>_d +3
			\end{array}\right)}_{\in \mathbb N\cap \left[ \ovun 1, \ovun 3 \right] }
		\right)
	\end{equation}
	
	$\longrightarrow$
	\begin{equation}
		\underline {\alpha \in \{1,...,3^d\}}
	\end{equation}
	Hence, under the condition eq. \ref{eq:initialCell:ParticleDistribution} no particle leaves the storage compartments of its process through the function $dist_{\left(\tilde{g},\mathcal{G}^1\right)}^{ALL}$. 
\end{proof}

\begin{proof}\textbf{of lemma \ref{lemma:distPlacesParticlesOnlyInsideDomain}}\\
	We need to prove that
	the function $dist_{\left(\tilde{g},\mathcal{G}^1\right)}^{ALL}$ places particles only inside storage compartments which represent cells inside the domain. Hence, processes at the domain's border do not have particles in their outer storage compartments.
	Be
	\begin{equation}
		\underline{w}=\left(w_1,...,w_d\right)^{\mathbf{T}}:=  {}^{\ovun I} \iota ( w)
	\end{equation}
	\begin{equation}
		\underline{\alpha}=\left(\alpha_1,...,\alpha_d\right)^{\mathbf{T}}:= \left\lfloor \tfrac{1}{r_{c}}(\underline{x}_j^2 -\underline{D}_{min})\right\rfloor - \underline w +\ovun 3,
	\end{equation}
	
	then
	\begin{multline}
		\forall w\in \{1,...,N_{cell}\} \ \forall p^2_j:=  \left<\left<\left< 
		step_{\mathcal{G}^1}^{ALL}\left(copy_{\tilde{g}}^{ALL}(\mathcal{P}^1)\right)
		\right>_w\right>_{\tfrac{3^d+1}{2}}\right>_j:\\ 
		(k\in \{1,...,d\} \land w_k=1) \ \rightarrow \ \alpha_k\in \{2,3\}\
		\land (k\in \{1,...,d\} \land w_k=I_k)\ \rightarrow \ \alpha_k\in \{1,2\}.
	\end{multline}

	We do a proof by contradiction.
	First, we prove the statement with $w_k=1$.
	Amusing
	\begin{equation}
		\exists k\in \{1,...,d\}: w_k=1\ \land \alpha_k=1
	\end{equation}
	
	then
	\begin{equation}
		\alpha_k=1=\left\lfloor \tfrac{1}{r_{c}}({}_{k}{x}_j^2 - {}_{k}{D}_{min})\right\rfloor - 1 + 3.
	\end{equation}
	
	$\longrightarrow$
	\begin{equation}
		-1=\left\lfloor \tfrac{1}{r_{c}}({}_{k}{x}_j^2 - {}_{k}{D}_{min})\right\rfloor
	\end{equation}
	
	$\longrightarrow$
	\begin{equation}
		\tfrac{1}{r_{c}}({}_{k}{x}_j^2 - {}_{k}{D}_{min})=\left[-1, 0  \right) 
	\end{equation}
	
	$\longrightarrow$
	\begin{equation}
		{}_{k}{x}_j^2=\left[-r_c+{}_{k}{D}_{min}, 0  \right) \quad \lightning
	\end{equation}
	This contradicts the condition that no particle leaves the domain (eq. \ref{eq:condition:particleStayInDomain}). Hence, in this case $\alpha_k \in \{2,3\}$.
	Now we prove the second statement with $w_k=I_k$.
	Amusing
	\begin{equation}
		\exists k\in \{1,...,d\}: w_k=I_k\ \land \alpha_k=3
	\end{equation}
	
	then
	\begin{equation}
		\alpha_k=3=\left\lfloor \tfrac{1}{r_{c}}({}_{k}{x}_j^2 - {}_{k}{D}_{min})\right\rfloor - I_k + 3.
	\end{equation}
	
	$\longrightarrow$
	\begin{equation}
		I_k=\left\lfloor \tfrac{1}{r_{c}}({}_{k}{x}_j^2 - {}_{k}{D}_{min})\right\rfloor
	\end{equation}
	
	$\longrightarrow$
	\begin{equation}
		\tfrac{1}{r_{c}}({}_{k}{x}_j^2 - {}_{k}{D}_{min})=\left[I_k, I_k+1  \right) 
	\end{equation}
	
	$\longrightarrow$
	\begin{equation}
		{}_{k}{x}_j^2=\left[I_k r_c+{}_{k}{D}_{min}, I_k r_c+{}_{k}{D}_{min}+r_c  \right) \quad \lightning
	\end{equation}
	
	$\xrightarrow{def.\ \ovun I\ (eq.\ \ref{eq:definitionOfI})}$
	\begin{equation}
		{}_{k}{x}_j^2>{}_{k}{D}_{max}) \quad \lightning
	\end{equation}
	This contradicts the condition that no particle leaves the domain (eq. \ref{eq:condition:particleStayInDomain}). Hence, in this case $\alpha_k \in \{1,2\}$.
	Taking both cases together, the function $dist_{\left(\tilde{g},\mathcal{G}^1\right)}^{ALL}$ places particles only inside storage compartments which represent cells inside the domain.
\end{proof}
\newpage

\begin{proof}\textbf{of lemma \ref{lemma:collectDistRestoreInitialDistributionCondition}}\\
	$\overline{\mathcal{P}}^1 :=step_{\mathcal{G}^1}^{ALL}\left(copy_{\tilde{g}}^{ALL}(\mathcal{P}^1)\right)$ does not necessarily fulfill the condition (eq. \ref{eq:initialCell:ParticleDistribution}) to reiterate  $step_{\mathcal{G}^2}^{ALL}\left(copy_{\tilde{g}}^{ALL}\left(\overline{\mathcal{P}}^1\right)\right)$. 
	The function $interact_g$ could miss interaction due to incomplete neighborhoods. Therefore, the particles in $\overline{\mathcal{P}}^1$ need to be redistributed such that the requirement (eq. \ref{eq:initialCell:ParticleDistribution}) is fulfilled. 
	The decomposition of the initial permutated particle tuple is stored in the center storage compartment of the processes. Hence, the condition (eq. \ref{eq:initialCell:ParticleDistribution}) can be rewritten for all state transition steps as 
	\begin{equation}
		\forall p^t_j\in 
		\underset{v=1}{\overset{N_{cell}}{\fullmoon }}\ \left<\left< \mathcal{P}^t\right>_v\right> _{\tfrac{3^d+1}{2}}: 
		p^t_j \in \left<\left< \mathcal{P}^t\right>_w\right> _{\tfrac{3^d+1}{2}},
	\end{equation}
	where 
	\begin{equation}
		w= {}^{\ovun{I}}\iota^{-1}\left(\left \lfloor \frac{1}{r_c} ( \underline{x}^t_j-\underline{D}_{min})\right \rfloor + \ovun{1} \right).
	\end{equation}
	We need to prove that the two functions $dist_{\left(\tilde{g},\mathcal{G}^1\right)}^{ALL}$ and $collect_{\tilde{g}}^{ALL}$ achieve this.
	\\[12pt]
	We prove this by induction. 
	Regarding the base case,
	we know that the condition  (eqs. \ref{eq:particleCellPlacementCondition}, \ref{eq:particleCellPlacementCondition:w}) is true for $\mathcal{P}^1$ by definition (eq. \ref{eq:initialCell:ParticleDistribution}).
	
	Regarding the induction step, we need to prove that 
	if $\mathcal P^t$ fulfills the condition (eqs. \ref{eq:particleCellPlacementCondition}, \ref{eq:particleCellPlacementCondition:w})
	then $\mathcal{P}^{t+1}=collect_{\tilde{g}}^{ALL}\left( dist_{\left(\tilde{g} ,\mathcal{G}^1\right)}^{ALL}\left( step_{\mathcal{G}^1}^{ALL}\left( copy_{\tilde{g}}^{ALL} (\mathcal{P}^t )\right)\right)\right)$ fulfills also the condition.

	$dist_{\left(\tilde{g},\mathcal{G}^1\right)}^{ALL}$ (eq. \ref{eq:distAll:distAll}) empties all storage compartments except the center storage compartment and then distributes all particles of the center storage compartments according to their position to the other storage compartments of each process. $\alpha$ gives the index of the new storage compartment for each particle. 
	It is trivial that it considers all particles and all processes since it simply iterates over them.
	
	$collect_{\tilde{g}}^{ALL}$ (eq. \ref{eq:collectAll:collectAll}) copies for each process its particles from the other processes corresponding storage compartments. It uses the same strategy as the function $copy_{\tilde{g}}^{ALL}$ to avoid overlapping conditions, hence, lemma \ref{lemma:copyAll:NoReadConflicts} is also valid for the $collect_{\tilde{g}}^{ALL}$ function. 
	The $collect_{\tilde{g}}^{ALL}$ function takes the particles from the ${}^{\ovun 3} \iota^{-1}\left(\ovun 4 -{}^{\ovun 3} \iota ( l)\right) $-th storage compartment of the $\beta(w,l)$-th process and stores it in the center storage compartment of the $w$-th process.

	We need to prove that a particle from each storage compartment of all processes ends up in the suitable process's central storage compartment according to the condition (eqs. \ref{eq:particleCellPlacementCondition}, \ref{eq:particleCellPlacementCondition:w}).
	We choose without restricting generality a particle $p$ from the ${}^{\ovun 3} \iota^{-1}\left(\ovun 4 -{}^{\ovun 3} \iota ( l)\right) $-th storage compartment of the $\beta(w,l)$-th process. Hence,
	\begin{equation}
		p=(...,x,...) \in \left<\left<step_{\mathcal{G}^1}^{ALL}\left( copy_{\tilde{g}}^{ALL} (\mathcal{P}^t )\right)\right>_{\beta(w,l)}\right>_{{}^{\ovun 3} \iota^{-1}\left(\ovun 4 -{}^{\ovun 3} \iota ( l)\right)}.
	\end{equation}
	
	Combining this with the distribution of the $ dist_{\left(\tilde{g} ,\mathcal{G}^1\right)}^{ALL}$ function  we get
	\begin{equation}
		\underbrace{\alpha}_{ \overset{!}{=} {}^{\ovun 3} \iota^{-1}\left(\ovun 4 -{}^{\ovun 3} \iota ( l)\right)} :=  {}^{\ovun 3} \iota^{-1}\left(\left\lfloor \tfrac{1}{r_{c}}(\underline{x} -\underline{D}_{min})\right\rfloor - {}^{\ovun I} \iota ( \underbrace{j}_{\overset{!}{=}\beta(w,l)}) +\ovun 3 \right),
	\end{equation}
	where $\alpha$ is set to ${}^{\ovun 3} \iota^{-1}\left(\ovun 4 -{}^{\ovun 3} \iota ( l)\right)$ the index of the storage compartment from that the particles are collected, and $j$ is set to $\beta(w,l)$ the index of the process from which is collected. Hence, our particle is on that storage compartment of that process and gets distributed by the mechanism of the $ dist_{\left(\tilde{g},\mathcal{G}^1\right)}^{ALL}$ function. We put the indices in and transform  
	\begin{equation}
		{}^{\ovun 3} \iota^{-1}\left(\ovun 4 -{}^{\ovun 3} \iota ( l)\right) =  {}^{\ovun 3} \iota^{-1}\left(\left\lfloor \tfrac{1}{r_{c}}(\underline{x} -\underline{D}_{min})\right\rfloor - {}^{\ovun I} \iota ( \beta(w,l)) +\ovun 3 \right),
	\end{equation}
	
	$\xrightarrow{lemma\ \ref{lemma:indexTransformIsBijectiv}}$
	\begin{equation}
		\ovun 4 -{}^{\ovun 3} \iota ( l) =  \left\lfloor \tfrac{1}{r_{c}}(\underline{x} -\underline{D}_{min})\right\rfloor - {}^{\ovun I} \iota ( \beta(w,l)) +\ovun 3,
	\end{equation}
	
	$\xrightarrow{def.\ of\  \beta \ (eq.\ \ref{eq:definition:beta})}$
	\begin{equation}
		\ovun 4 -{}^{\ovun 3} \iota ( l) =  \left\lfloor \tfrac{1}{r_{c}}(\underline{x} -\underline{D}_{min})\right\rfloor - {}^{\ovun I} \iota \left( {}^{\ovun I} \iota ^{-1}\left(  {}^{\ovun{I}} \iota (w)+{}^{\ovun{3}} \iota (l)-\ovun{2}\right)\right) +\ovun 3,
	\end{equation}
	
	$\xrightarrow{lemma\ \ref{lemma:indexTransformIsBijectiv}}$
	\begin{equation}
		\ovun 4 -{}^{\ovun 3} \iota ( l) =  \left\lfloor \tfrac{1}{r_{c}}(\underline{x} -\underline{D}_{min})\right\rfloor -   {}^{\ovun{I}} \iota (w)-{}^{\ovun{3}} \iota (l)+\ovun{2} +\ovun 3,
	\end{equation}
	
	$\longrightarrow$
	\begin{equation}
		{}^{\ovun{I}} \iota (w)   =  \left\lfloor \tfrac{1}{r_{c}}(\underline{x} -\underline{D}_{min})\right\rfloor +\ovun 1,
	\end{equation}
	
	$\xrightarrow{lemma\ \ref{lemma:indexTransformIsBijectiv}}$
	\begin{equation}
		w  =  {}^{\ovun{I}}\iota^{-1} \left(\left\lfloor \tfrac{1}{r_{c}}(\underline{x} -\underline{D}_{min})\right\rfloor +\ovun 1\right),
	\end{equation}
	The functions $ dist_{\left(\tilde{g} ,\mathcal{G}^1\right)}^{ALL}$ 
	and $collect_{\tilde{g}}^{ALL}$ put our particle $p$ in the central storage compartment of the $w$-th process
	\begin{equation}
		p\in \left<\left<collect_{\tilde{g}}^{ALL}\left( dist_{\left(\tilde{g} ,\mathcal{G}^1\right)}^{ALL}\left( step_{\mathcal{G}^1}^{ALL}\left( copy_{\tilde{g}}^{ALL} (\mathcal{P}^t )\right)\right)\right)\right>_w\right> _{\tfrac{3^d+1}{2}}.
	\end{equation}
	Hence, 
	\begin{equation}
		p\in \left<\left<\mathcal{P}^{t+1} \right>_w\right> _{\tfrac{3^d+1}{2}}.
	\end{equation}
	fulfills the condition (eqs. \ref{eq:particleCellPlacementCondition}, \ref{eq:particleCellPlacementCondition:w}).
\end{proof}


\begin{thebibliography}{24}	
	\ifx \showCODEN    \undefined \def \showCODEN     #1{\unskip}     \fi
	\ifx \showDOI      \undefined \def \showDOI       #1{#1}\fi
	\ifx \showISBNx    \undefined \def \showISBNx     #1{\unskip}     \fi
	\ifx \showISBNxiii \undefined \def \showISBNxiii  #1{\unskip}     \fi
	\ifx \showISSN     \undefined \def \showISSN      #1{\unskip}     \fi
	\ifx \showLCCN     \undefined \def \showLCCN      #1{\unskip}     \fi
	\ifx \shownote     \undefined \def \shownote      #1{#1}          \fi
	\ifx \showarticletitle \undefined \def \showarticletitle #1{#1}   \fi
	\ifx \showURL      \undefined \def \showURL       {\relax}        \fi

	\providecommand\bibfield[2]{#2}
	\providecommand\bibinfo[2]{#2}
	\providecommand\natexlab[1]{#1}
	\providecommand\showeprint[2][]{arXiv:#2}
	
	\bibitem
	{Afshar:2016}
	\bibfield{author}{\bibinfo{person}{Yaser Afshar} {and} \bibinfo{person}{Ivo~F.
			Sbalzarini}.} \bibinfo{year}{2016}\natexlab{}.
	\newblock \showarticletitle{A Parallel Distributed-Memory Particle Method
		Enables Acquisition-Rate Segmentation of Large Fluorescence Microscopy
		Images}.
	\newblock \bibinfo{journal}{\emph{PLoS One}} \bibinfo{volume}{11},
	\bibinfo{number}{4} (\bibinfo{year}{2016}), \bibinfo{pages}{e0152528}.
	\newblock
	\urldef\tempurl
	\url{https://doi.org/10.1371/journal.pone.0152528}
	\showDOI{\tempurl}
	
	
	\bibitem
	{Alder:1957}
	\bibfield{author}{\bibinfo{person}{B~J Alder} {and} \bibinfo{person}{T~E
			Wainwright}.} \bibinfo{year}{1957}\natexlab{}.
	\newblock \showarticletitle{Molecular dynamics simulation of hard sphere
		system}.
	\newblock \bibinfo{journal}{\emph{J.\ Chem.\ Phys.}}  \bibinfo{volume}{27}
	(\bibinfo{year}{1957}), \bibinfo{pages}{1208--1218}.
	\newblock
	
	
	\bibitem
	{Amdahl1967}
	\bibfield{author}{\bibinfo{person}{Gene~M Amdahl}.}
	\bibinfo{year}{1967}\natexlab{}.
	\newblock \showarticletitle{Validity of the single processor approach to
		achieving large scale computing capabilities}. In
	\bibinfo{booktitle}{\emph{Proceedings of the April 18-20, 1967, spring joint
			computer conference}}. \bibinfo{pages}{483--485}.
	\newblock
	
	
	\bibitem
	{Bamme2021}
	\bibfield{author}{\bibinfo{person}{Johannes Bamme} {and}
		\bibinfo{person}{Ivo~F. Sbalzarini}.} \bibinfo{year}{2021}\natexlab{}.
	\newblock \showarticletitle{A Mathematical Definition of Particle Methods}.
	\newblock \bibinfo{journal}{\emph{CoRR}}  \bibinfo{volume}{abs/2105.05637}
	(\bibinfo{year}{2021}).
	\newblock
	\urldef\tempurl
	\url{https://doi.org/10.48550/ARXIV.2105.05637}
	\showDOI{\tempurl}
	\showeprint[arXiv]{2105.05637}
	
	
	\bibitem
	{Bourantas:2016}
	\bibfield{author}{\bibinfo{person}{George~C. Bourantas},
		\bibinfo{person}{Bevan~L. Cheeseman}, \bibinfo{person}{Rajesh Ramaswamy},
		{and} \bibinfo{person}{Ivo~F. Sbalzarini}.} \bibinfo{year}{2016}\natexlab{}.
	\newblock \showarticletitle{Using {DC PSE} operator discretization in
		{E}ulerian meshless collocation methods improves their robustness in complex
		geometries}.
	\newblock \bibinfo{journal}{\emph{Computers \& Fluids}}  \bibinfo{volume}{136}
	(\bibinfo{year}{2016}), \bibinfo{pages}{285--300}.
	\newblock
	
	
	\bibitem
	{Cardinale:2012}
	\bibfield{author}{\bibinfo{person}{Janick Cardinale},
		\bibinfo{person}{Gr\'{e}gory Paul}, {and} \bibinfo{person}{Ivo~F.
			Sbalzarini}.} \bibinfo{year}{2012}\natexlab{}.
	\newblock \showarticletitle{Discrete region competition for unknown numbers of
		connected regions}.
	\newblock \bibinfo{journal}{\emph{IEEE Trans. Image Process.}}
	\bibinfo{volume}{21}, \bibinfo{number}{8} (\bibinfo{year}{2012}),
	\bibinfo{pages}{3531--3545}.
	\newblock
	
	
	\bibitem
	{Cottet:1990}
	\bibfield{author}{\bibinfo{person}{G.~H. Cottet} {and} \bibinfo{person}{S.
			Mas-Gallic}.} \bibinfo{year}{1990}\natexlab{}.
	\newblock \showarticletitle{A Particle Method to Solve the {N}avier-{S}tokes
		System}.
	\newblock \bibinfo{journal}{\emph{Numer.\ Math.}}  \bibinfo{volume}{57}
	(\bibinfo{year}{1990}), \bibinfo{pages}{805--827}.
	\newblock
	
	
	\bibitem
	{Degond:1989a}
	\bibfield{author}{\bibinfo{person}{P. Degond} {and} \bibinfo{person}{S.
			Mas-Gallic}.} \bibinfo{year}{1989}\natexlab{}.
	\newblock \showarticletitle{The Weighted Particle Method for
		Convection-Diffusion Equations. {P}art 1: {T}he Case of an Isotropic
		Viscosity}.
	\newblock \bibinfo{journal}{\emph{Math.\ Comput.}} \bibinfo{volume}{53},
	\bibinfo{number}{188} (\bibinfo{year}{1989}), \bibinfo{pages}{485--507}.
	\newblock
	
	
	\bibitem
	{Eldredge:2002}
	\bibfield{author}{\bibinfo{person}{Jeff~D. Eldredge}, \bibinfo{person}{Anthony
			Leonard}, {and} \bibinfo{person}{Tim Colonius}.}
	\bibinfo{year}{2002}\natexlab{}.
	\newblock \showarticletitle{A General Deterministic Treatment of Derivatives in
		Particle Methods}.
	\newblock \bibinfo{journal}{\emph{J.\ Comput.\ Phys.}}  \bibinfo{volume}{180}
	(\bibinfo{year}{2002}), \bibinfo{pages}{686--709}.
	\newblock
	
	
	\bibitem
	{Gingold:1977}
	\bibfield{author}{\bibinfo{person}{R.~A. Gingold} {and} \bibinfo{person}{J.~J.
			Monaghan}.} \bibinfo{year}{1977}\natexlab{}.
	\newblock \showarticletitle{Smoothed particle hydrodynamics - Theory and
		application to non-spherical stars}.
	\newblock \bibinfo{journal}{\emph{Royal Astronomical Society, Montly Notices}}
	\bibinfo{volume}{181} (\bibinfo{year}{1977}), \bibinfo{pages}{375--378}.
	\newblock
	
	
	\bibitem
	{Gustafson1988}
	\bibfield{author}{\bibinfo{person}{John Gustafson}.}
	\bibinfo{year}{1988}\natexlab{}.
	\newblock \showarticletitle{Reevaluating Amdahl's Law}.
	\newblock \bibinfo{journal}{\emph{Commun. ACM}}  \bibinfo{volume}{31}
	(\bibinfo{date}{05} \bibinfo{year}{1988}), \bibinfo{pages}{532--533}.
	\newblock
	\urldef\tempurl
	\url{https://doi.org/10.1145/42411.42415}
	\showDOI{\tempurl}
	
	
	\bibitem
	{Hockney:1988}
	\bibfield{author}{\bibinfo{person}{R.~W. Hockney} {and} \bibinfo{person}{J.~W.
			Eastwood}.} \bibinfo{year}{1988}\natexlab{}.
	\newblock \bibinfo{booktitle}{\emph{Computer Simulation using Particles}}.
	\newblock \bibinfo{publisher}{Institute of Physics Publishing}.
	\newblock
	
	
	\bibitem
	{Incardona:2019}
	\bibfield{author}{\bibinfo{person}{Pietro Incardona}, \bibinfo{person}{Antonio
			Leo}, \bibinfo{person}{Yaroslav Zaluzhnyi}, \bibinfo{person}{Rajesh
			Ramaswamy}, {and} \bibinfo{person}{Ivo~F. Sbalzarini}.}
	\bibinfo{year}{2019}\natexlab{}.
	\newblock \showarticletitle{{OpenFPM}: A scalable open framework for particle
		and particle-mesh codes on parallel computers}.
	\newblock \bibinfo{journal}{\emph{Comput.\ Phys.\ Commun.}}
	\bibinfo{volume}{241} (\bibinfo{year}{2019}), \bibinfo{pages}{155--177}.
	\newblock
	
	
	\bibitem
	{Iwasawa:2016}
	\bibfield{author}{\bibinfo{person}{Masaki Iwasawa}, \bibinfo{person}{Ataru
			Tanikawa}, \bibinfo{person}{Natsuki Hosono}, \bibinfo{person}{Keigo
			Nitadori}, \bibinfo{person}{Takayuki Muranushi}, {and}
		\bibinfo{person}{Junichiro Makino}.} \bibinfo{year}{2016}\natexlab{}.
	\newblock \showarticletitle{Implementation and performance of {FDPS}: a
		framework for developing parallel particle simulation codes}.
	\newblock \bibinfo{journal}{\emph{Publications of the Astronomical Society of
			Japan}} \bibinfo{volume}{68}, \bibinfo{number}{4} (\bibinfo{year}{2016}),
	\bibinfo{pages}{54}.
	\newblock
	
	
	\bibitem
	{Karol:2018}
	\bibfield{author}{\bibinfo{person}{Sven Karol}, \bibinfo{person}{Tobias Nett},
		\bibinfo{person}{Jeronimo Castrillon}, {and} \bibinfo{person}{Ivo~F.
			Sbalzarini}.} \bibinfo{year}{2018}\natexlab{}.
	\newblock \showarticletitle{A Domain-Specific Language and Editor for Parallel
		Particle Methods}.
	\newblock \bibinfo{journal}{\emph{ACM Trans. Math. Softw.}}
	\bibinfo{volume}{44}, \bibinfo{number}{3} (\bibinfo{year}{2018}),
	\bibinfo{pages}{34}.
	\newblock
	
	
	\bibitem
	{Karr2012}
	\bibfield{author}{\bibinfo{person}{Jonathan~R. Karr},
		\bibinfo{person}{Jayodita~C. Sanghvi}, \bibinfo{person}{Derek~N. Macklin},
		\bibinfo{person}{Miriam~V. Gutschow}, \bibinfo{person}{Jared~M. Jacobs},
		\bibinfo{person}{Benjamin Bolival}, \bibinfo{person}{Nacyra Assad-Garcia},
		\bibinfo{person}{John~I. Glass}, {and} \bibinfo{person}{Markus~W. Covert}.}
	\bibinfo{year}{2012}\natexlab{}.
	\newblock \showarticletitle{A Whole-Cell Computational Model Predicts Phenotype
		from Genotype}.
	\newblock \bibinfo{journal}{\emph{Cell}}  \bibinfo{volume}{150}
	(\bibinfo{year}{2012}), \bibinfo{pages}{389--401}.
	\newblock
	
	
	\bibitem
	{Liu:1995}
	\bibfield{author}{\bibinfo{person}{Wing~Kam Liu}, \bibinfo{person}{Sukky Jun},
		{and} \bibinfo{person}{Yi~Fei Zhang}.} \bibinfo{year}{1995}\natexlab{}.
	\newblock \showarticletitle{Reproducing Kernel Particle Methods}.
	\newblock \bibinfo{journal}{\emph{Int.\ J.\ Numer.\ Meth.\ Fluids}}
	\bibinfo{volume}{20} (\bibinfo{year}{1995}), \bibinfo{pages}{1081--1106}.
	\newblock
	
	
	\bibitem
	{Monaghan:2005}
	\bibfield{author}{\bibinfo{person}{J.~J. Monaghan}.}
	\bibinfo{year}{2005}\natexlab{}.
	\newblock \showarticletitle{Smoothed particle hydrodynamics}.
	\newblock \bibinfo{journal}{\emph{Rep.\ Prog.\ Phys.}}  \bibinfo{volume}{68}
	(\bibinfo{year}{2005}), \bibinfo{pages}{1703--1759}.
	\newblock
	
	
	\bibitem
	{Nelson1996}
	\bibfield{author}{\bibinfo{person}{Mark~T. Nelson}, \bibinfo{person}{William
			Humphrey}, \bibinfo{person}{Attila Gursoy}, \bibinfo{person}{Andrew Dalke},
		\bibinfo{person}{Laxmikant~V. Kalé}, \bibinfo{person}{Robert~D. Skeel},
		{and} \bibinfo{person}{Klaus Schulten}.} \bibinfo{year}{1996}\natexlab{}.
	\newblock \showarticletitle{NAMD: a Parallel, Object-Oriented Molecular
		Dynamics Program}.
	\newblock \bibinfo{journal}{\emph{The International Journal of Supercomputer
			Applications and High Performance Computing}} \bibinfo{volume}{10},
	\bibinfo{number}{4} (\bibinfo{year}{1996}), \bibinfo{pages}{251--268}.
	\newblock
	\urldef\tempurl
	\url{https://doi.org/10.1177/109434209601000401}
	\showDOI{\tempurl}
	\showeprint{https://doi.org/10.1177/109434209601000401}
	
	
	\bibitem
	{Reynders:1996a}
	\bibfield{author}{\bibinfo{person}{J.V.W. Reynders}, \bibinfo{person}{J.C.
			Cummings}, \bibinfo{person}{M. Tholburn}, \bibinfo{person}{P.J. Hinker},
		\bibinfo{person}{S.R. Atlas}, \bibinfo{person}{S. Banerjee},
		\bibinfo{person}{M. Srikant}, \bibinfo{person}{W.F. Humphrey},
		\bibinfo{person}{S.R. Karmesin}, {and} \bibinfo{person}{K. Keahey}.}
	\bibinfo{year}{1996}\natexlab{}.
	\newblock \showarticletitle{POOMA: a framework for scientific simulation on
		parallel architectures}. In \bibinfo{booktitle}{\emph{Proceedings. First
			International Workshop on High-Level Programming Models and Supportive
			Environments}}, \bibfield{editor}{\bibinfo{person}{A.~Bode},
		\bibinfo{person}{M.~Gerndt}, \bibinfo{person}{R.G. Hackenberg}, {and}
		\bibinfo{person}{H.~Hellwagner}} (Eds.). Tech. Univ. Munchen; Res. Centre
	Julich; Central Inst. Appl. Math.; 10th IEEE Int. Parallel Process.
	Symposium; IEEE Comput. Soc. Tech. Committee on Parallel Process.; ACM
	SIGARCH, \bibinfo{publisher}{{IEEE Comput. Soc. Press}},
	\bibinfo{address}{Los Alamitos, CA, USA}, \bibinfo{pages}{41--49}.
	\newblock
	\showISBNx{{0 8186 7567 5}}
	
	
	\bibitem
	{Sbalzarini:2006b}
	\bibfield{author}{\bibinfo{person}{I.~F. Sbalzarini}, \bibinfo{person}{J.~H.
			Walther}, \bibinfo{person}{M. Bergdorf}, \bibinfo{person}{S.~E. Hieber},
		\bibinfo{person}{E.~M. Kotsalis}, {and} \bibinfo{person}{P. Koumoutsakos}.}
	\bibinfo{year}{2006}\natexlab{}.
	\newblock \showarticletitle{{PPM} -- A Highly Efficient Parallel Particle-Mesh
		Library for the Simulation of Continuum Systems}.
	\newblock \bibinfo{journal}{\emph{J.\ Comput.\ Phys.}} \bibinfo{volume}{215},
	\bibinfo{number}{2} (\bibinfo{year}{2006}), \bibinfo{pages}{566--588}.
	\newblock
	
	
	\bibitem
	{Schrader:2010}
	\bibfield{author}{\bibinfo{person}{Birte Schrader}, \bibinfo{person}{Sylvain
			Reboux}, {and} \bibinfo{person}{Ivo~F. Sbalzarini}.}
	\bibinfo{year}{2010}\natexlab{}.
	\newblock \showarticletitle{Discretization Correction of General Integral {PSE}
		Operators in Particle Methods}.
	\newblock \bibinfo{journal}{\emph{J.\ Comput.\ Phys.}}  \bibinfo{volume}{229}
	(\bibinfo{year}{2010}), \bibinfo{pages}{4159--4182}.
	\newblock
	
	
	\bibitem
	{Verma2018}
	\bibfield{author}{\bibinfo{person}{Siddhartha Verma}, \bibinfo{person}{Guido
			Novati}, {and} \bibinfo{person}{Petros Koumoutsakos}.}
	\bibinfo{year}{2018}\natexlab{}.
	\newblock \showarticletitle{Efficient collective swimming by harnessing
		vortices through deep reinforcement learning}.
	\newblock \bibinfo{journal}{\emph{Proceedings of the National Academy of
			Sciences}} \bibinfo{volume}{115}, \bibinfo{number}{23}
	(\bibinfo{year}{2018}), \bibinfo{pages}{5849--5854}.
	\newblock
	\urldef\tempurl
	\url{https://doi.org/10.1073/pnas.1800923115}
	\showDOI{\tempurl}
	\showeprint{https://www.pnas.org/doi/pdf/10.1073/pnas.1800923115}
	
	
	\bibitem
	{Walther:2009}
	\bibfield{author}{\bibinfo{person}{Jens~H. Walther} {and}
		\bibinfo{person}{Ivo~F. Sbalzarini}.} \bibinfo{year}{2009}\natexlab{}.
	\newblock \showarticletitle{Large-scale parallel discrete element simulations
		of granular flow}.
	\newblock \bibinfo{journal}{\emph{Engineering Computations}}
	\bibinfo{volume}{26}, \bibinfo{number}{6} (\bibinfo{year}{2009}),
	\bibinfo{pages}{688--697}.
	\newblock
	
	
\end{thebibliography}
\end{document}